\documentclass{article}

\usepackage{arxiv}
\usepackage[utf8]{inputenc} 
\usepackage[T1]{fontenc}    
\usepackage{hyperref}       
\usepackage{url}            
\usepackage{booktabs}       
\usepackage{amsmath,amssymb,amsfonts}%
\usepackage{amsthm}%
\usepackage{mathrsfs}%
\usepackage{nicefrac}       
\usepackage{microtype}      
\usepackage{graphicx}
\usepackage[sort,numbers]{natbib}
\usepackage{doi}

\usepackage{float}
\usepackage{longtable}
\usepackage[font=footnotesize,labelfont=bf]{caption}
\usepackage{subcaption}
\usepackage{makecell, rotating}
\usepackage{colortbl}
\usepackage[capitalise,nameinlink]{cleveref}
\usepackage{array}
\usepackage{ragged2e}
\usepackage{tabularray}
\usepackage{multirow}%
\usepackage{authblk}
\usepackage{setspace}
\usepackage{pdflscape}

\newcommand{\mt}[1]{\mathbf{#1}} 
\newcommand{\lb}[1]{\left( #1 \right)}

\newcommand{\jdist}[2]{\text{#1}\left( #2 \right)}
\newcommand{\jdistu}[3]{\text{#1}_{#2}\left( #3 \right)}

\newcommand{\etal}{\emph{et al.} }
\NewDocumentCommand{\rot}{O{90} O{1em} m}{\makebox[#2][l]{\rotatebox{#1}{#3}}}%

\title{Creating area level indices of behaviours impacting cancer in Australia with a Bayesian generalised shared component model}

\author{
\href{https://orcid.org/0000-0002-4666-5900}{\includegraphics[scale=0.06]{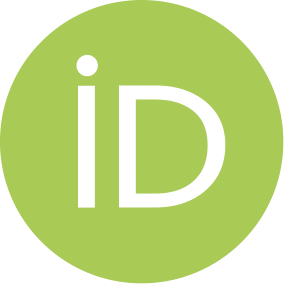}\hspace{1mm}James Hogg$^1$}\thanks{Corresponding author: \href{mailto:james.hogg@hdr.qut.edu.au}{james.hogg@hdr.qut.edu.au}} 
\hspace{5mm}%
\href{https://orcid.org/0000-0001-9041-9531}{\includegraphics[scale=0.06]{plots/orcid.png}\hspace{1mm}Susanna Cramb$^3$} 
\hspace{5mm}%
\href{https://orcid.org/0000-0002-8161-0358}{\includegraphics[scale=0.06]{plots/orcid.png}\hspace{1mm}Jessica Cameron$^{1,2}$} 
\hspace{5mm}%
\href{https://orcid.org/0000-0001-8576-8868}{\includegraphics[scale=0.06]{plots/orcid.png}\hspace{1mm}Peter Baade$^{1,2}$} 
\hspace{5mm}%
\href{https://orcid.org/0000-0001-8625-9168}{\includegraphics[scale=0.06]{plots/orcid.png}\hspace{1mm}Kerrie Mengersen$^{1}$}}
\date{%
    $^1$Centre for Data Science, Queensland University of Technology (QUT), 2 George St, Brisbane City, 4000, Queensland, Australia\\%
    $^2$Viertel Cancer Research Centre, Cancer Council Queensland (CCQ), 553 Gregory Terrace, Fortitude Valley, 4006, Queensland, Australia\\%
    $^3$Australian Centre for Health Services Innovation, School of Public Health and Social Work, Queensland University of Technology (QUT), 2 George St, Brisbane City, 4000, Queensland, Australia\\[2ex]%
}

\renewcommand{\headeright}{}
\renewcommand{\undertitle}{}
\renewcommand{\shorttitle}{Bayesian GSCM}

\hypersetup{
colorlinks,
allcolors=blue,
pdftitle={Creating area level indices of behaviours impacting cancer in Australia with a Bayesian generalised shared component model},
pdfsubject={},
pdfauthor={Hogg, Cramb, Cameron, Baade, Mengersen},
pdfkeywords={Bayesian statistics, latent models, spatial statistics, unhealthy behaviors, index},
}

\begin{document}
\maketitle

\begin{abstract}
This study develops a model-based index creation approach called the Generalised Shared Component Model (GSCM) by drawing on the large field of factor models. The proposed fully Bayesian approach accommodates heteroscedastic model error, multiple shared factors and flexible spatial priors. Moreover, unlike previous index approaches, our model provides indices with uncertainty. Focusing on unhealthy behaviors that lead to cancer, the proposed GSCM is used to develop the Area Indices of Behaviors Impacting Cancer product --- representing the first area level cancer risk factor index in Australia. This advancement aids in identifying communities with elevated cancer risk, facilitating targeted health interventions. 
\end{abstract}


\section{Introduction} \label{sec:introduction}

In the past two decades, the resolution and accessibility of health data for small geographic areas have improved considerably. While these improvements allow for more granular insights into spatial disparities in specific diseases or health characteristics, the increasing quantity and complexity of such data can make it more challenging to collate, draw meaningful conclusions and formulate effective policy decisions \cite{RN674}. Recently, interactive and publicly available health atlases (e.g., Australian Cancer Atlas \cite{RN26} or PLACES platform \cite{RN426}) have gained popularity as valuable tools for disseminating health data for small areas \cite{RN26, RN113, RN426}. These atlases often provide estimates and measures of uncertainty (e.g., measurement error) for multiple health measures, opening up avenues for further spatial analysis \cite{RN613, RN9}.

Multidimensional data often contain redundancy; that is, some dimensions may not provide substantive additional information. In this case, the data could be reasonably summarised by a lower-dimensional latent space, called an index \cite{RN674}. An index should measure a multifaceted concept that cannot be captured by a single feature. Indices often offer a simple interpretation of a complex phenomenon, can help determine policy priorities, and can be used to benchmark or monitor the performance of current policies \cite{RN360}. 

There are a plethora of methods routinely used to derive indices \cite{RN359, RN360} including rank sum, standardisation \cite{RN360} and principal component analysis (PCA) \cite{RN560}. However, these methods do not naturally accommodate spatial autocorrelation and area-specific uncertainty, which are common in atlas data. Since atlas data are primarily derived from models, their uncertainty is largely model-based. Like other work \cite{RN376}, we use the generic term measurement error to denote this model-based uncertainty. 

While spatial PCA can accommodate spatial autocorrelation \cite{RN664, Lopez2024}, model-based approaches are gaining popularity in the creation of indices \cite{RN674, RN671, RN670, RN687} as these approaches can also accommodate measurement error and provide uncertainty measures. Furthermore, previous work highlights that by fitting Bayesian models \cite{RN671, RN670, RN504}, researchers can provide probabilistic statements about index values. This facilitates a sophisticated decision-making process that uses the distributions of index values rather than point estimates. With access to distributions, researchers can more effectively determine whether variations in index values are truly meaningful. 

Shared component models (SCMs) \cite{RN611} have been used to create indices from spatial data \cite{RN670}. These models assume that the input data originate from a linear combination of a \emph{single} shared factor and residual errors \cite{RN618, RN659}. However, conventional SCMs do not naturally accommodate measurement error \cite{RN546, RN6, RN364, RN496,  RN614} and can be limiting when the input data lack strong correlations \cite{RN546}. 

This study introduces a Bayesian generalised SCM (GSCM) capable of accommodating heteroscedastic measurement error, multiple shared factors and flexible spatial priors \cite{RN8}. Innovations are introduced with respect to the modelling, computation, and application. In terms of the model, while Bayesian spatial factor models have been employed previously to generate health indices with uncertainty measures \cite{RN671, RN687}, the GSCM bridges the mathematical and intuitive gaps between SCMs and Bayesian spatial factor models (BSFM); a connection that, to our knowledge, has yet to be acknowledged. To help distinguish between SCMs and BSFMs, we define a SCM as a model with a single shared factor and spatial dependencies in both the shared and residual terms. A BSFM, on the other hand, has at least one shared factor and spatial dependency in at least one component of the model.  

On the computational front, the model was fitted using Stan \cite{RN452}, a flexible tool that facilitates efficient analysis of diverse Bayesian models. As a side effect of this work, we also present an efficient implementation of the Leroux prior \cite{RN366} in Stan, which, to our knowledge, represents the first feasible implementation of this increasingly common spatial prior using Hamiltonian Monte Carlo.

The motivation for this work is a significant public health challenge; cancer. In light of the substantial global health burden \cite{sung2021global}, cancer prevention strategies focus on reducing the prevalence of unhealthy behaviors (e.g., risk factors) that can lead to cancer such as drinking, smoking, obesity, poor diet, and inadequate activity \cite{RN165}. Previous research has found spatial disparities in unhealthy behaviors that can lead to cancer \cite{RN113, RN426, RN2, selfcite2}, such as smoking \cite{RN607}. Yet, while area-level health indices are commonplace \cite{RN358}, cancer-specific indices have primarily been developed at the individual level \cite{RN352, RN354}. Area-level cancer risk indices have remained unexplored. 

To address this gap and showcase the potential of the GSCM approach, we conduct a comprehensive case study using publicly available modelled data on unhealthy behaviors in Australia. The aim of this study is threefold:
\begin{itemize}
    \item[(1)] generate a summary measure of unhealthy behaviors in Australia to help understand any underlying spatial patterns;
    \item[(2)] identify areas with elevated prevalence of unhealthy behaviors that may cause cancer;
    \item[(3)] provide decision support to help policymakers focus their cancer prevention strategies.
\end{itemize}
To do this, we developed the first area-level cancer risk index in Australia called the Area Indices of Behaviors Impacting Cancer (AIBIC), which provides a relative summary measure to compare the prevalence of unhealthy behaviors between areas. Unlike other applications of model-based index creation, which provide a single index only \cite{RN671, RN687, RN670}, we provide four indices in the AIBIC and explore their distinct properties. The first two indices stem directly from the fitted GSCM, while the latter two are unique transformations of the former pair. 


\section{Methods}

\subsection{Data}

The presented analyses utilize data from our previous Bayesian spatial small area estimation study \cite{selfcite2} which used data from the 2017-2018 National Health Survey (NHS). The NHS, conducted by the Australian Bureau of Statistics (ABS), is an Australia-wide population-level health survey conducted every 3-4 years \cite{RN478, RN598}. In our previous study \cite{selfcite2} we generated modelled estimates and measures of uncertainty for the proportion of persons aged 15 years and older in 2017-18 engaging in behaviors that increase the risk of being diagnosed with cancer \cite{RN165}. Modelled results were provided for 2221 small areas, known as Statistical Area Level 2 or ``SA2s'', as defined by the 2016 Australian Statistical Geography Standard (ASGS) \cite{RN348}. 

For the AIBIC product, we used five features from Hogg \etal \cite{selfcite2}, namely, the proportion of the following unhealthy behaviors: current smokers, risky alcohol consumption, overweight/obese, inadequate diet and inadequate physical activity based on leisure only. The selection of these unhealthy behaviors followed the same rationale as Hogg \etal \cite{selfcite2}, involving consultations with experts, a review of relevant literature \cite{RN123,RN121,RN485,RN165} and the evaluation of data availability in the 2017-2018 NHS.

These data on unhealthy behaviors were available as point estimates and standard deviations of proportions. For all the unhealthy behaviors, each was defined so that a higher proportion indicated a higher prevalence of poorer health behavior \cite{selfcite2}. To enable the use of a Gaussian likelihood in our model we first applied the inverse logit transformation, resulting in $\mt{Y}_{k}$ and $\mt{S}_{k}$ being $N$-dimensional vectors of estimates and corresponding standard deviations for risk factor $k=1, \dots, 5$, where $N = 2221$ is the total number of small areas (i.e., SA2s). Finally, these logit-transformed point estimates were centered and scaled, with the corresponding standard deviations appropriately adjusted (see Section A of the Additional File). The final set of input features is available on Github \cite{hogg2024gscm}.

Alongside the data on unhealthy behaviors, in this study, we utilised population data for persons 15 years and older at the SA2 level. The vector of population counts, $\mt{P} = \lb{P_1, \dots, P_N}$, was sourced from the ABS's Estimated Resident Population product and calculated as the average annual count for 2017 and 2018 \cite{RN650}.

\subsubsection{Risk factors}

Current smoking was defined as those who reported to be current smokers (including daily, weekly or less than weekly), and had smoked at least 100 cigarettes in their lifetime.

Risky alcohol consumption was defined as individuals who exceeded the revised 2020 National Health and Medical Research Council (NHMRC) guidelines \cite{RN516} of up to 10 standard drinks/week and no more than 4 standard drinks in any day. Compliance with the guidelines was assessed using self-reported alcohol consumption during the last three drinking days from the preceding seven days.

Inadequate diet, which was based on self-reported diet, was defined as those who did not meet both the fruit (2 serves/day) and vegetable (5 serves/day) 2013 NHMRC Australian Dietary guidelines \cite{RN520}. Participants of the 2017-18 NHS were asked to report the number of servings of fruit and serves of vegetables they usually ate each day.

Overweight/obese was defined as persons with a Body Mass Index greater or equal to 25. The NHS data collects both measured and self-reported height and weight. Measurements were voluntary, so 40\% of the measured values in the NHS were imputed by the ABS using the hot-decking method \cite{RN508}. 

Inadequate physical activity based on leisure only was defined as those who did not meet the 2014 Department of Health Physical Activity guidelines \cite{RN517}. This guideline stipulated that each week, adults aged 18-64 should either do 2 1/2 to 5 hours of moderate-intensity physical activity or 1 1/4 to 2 1/2 hours of vigorous-intensity physical activity or an equivalent combination of both, plus muscle-strengthening activities at least 2 days each week.




\subsection{Statistical Model} \label{sec:gscm}

Shared component models (SCM) are a multivariate statistical tool developed in the context of disease mapping. The traditional assumption underpinning the SCM is that the input features share a single common factor \cite{RN546, RN6} or latent field that ``manifests'' in the separate features. To accommodate residual heterogeneity not captured by the shared factor, SCMs also include feature-specific residual errors \cite{RN611}. In disease mapping, these models assume spatial dependence in both the shared factor and the feature-specific residual errors. Borrowing from the massive literature on factor models, incorporating multiple factors into the SCM enhances its ability to explain variation in the multivariate data, improves the approximation of the covariance structure, and facilitates more informative inferences.

Given the expressed interest in accommodating the uncertainty of the input features, there is a need to include area-specific measurement error. Following the classical measurement error framework \cite{RN549}, which is abundant in spatial models \cite{RN106, RN108}, we adopt a similar method to that of Jahan \etal \cite{RN613}. With the constraint that the number of shared factors ($L$) was much fewer than the number of features (i.e., $L << K$), the proposed GSCM is written as a $N$-dimensional multivariate normal distribution for the $k$th feature, 

\begin{eqnarray}
    \mt{Y}_{k} & \sim & \jdistu{MVN}{N}{\boldsymbol{\mu}_{k}, \jdist{diag}{\boldsymbol{\sigma}_k}} \label{eq:gscm}
    \\
    \boldsymbol{\mu}_{k} & = & \mt{z} \lb{ \boldsymbol{\Lambda}_k }^T + \boldsymbol{\epsilon}_{k} \nonumber,
\end{eqnarray}

where $\mt{Y}_{k}$, $\boldsymbol{\sigma}_k$ and $\boldsymbol{\mu}_{k}$ are $N$-dimensional vectors of estimates, corresponding standard deviations and true values for the $k$th risk factor. Unlike Jahan \etal \cite{RN613}, who placed a prior on their standard deviation estimates, in this work $\boldsymbol{\sigma}_k$ is taken to be equal to $\mt{S}_{k}$ (i.e., the standard deviations from Hogg \etal \cite{selfcite2}).  

In line with the classical measurement error framework \cite{RN549}, while the estimates are assumed independent, the true values are assumed to exhibit a complex correlation structure, captured in the $\boldsymbol{\mu}_{k}$'s. $\mt{z}$ is a $N \times L$ matrix of shared factor scores, with independent columns and $\boldsymbol{\epsilon}_{k}$ a $N$-dimensional vector of feature-specific residual errors for the $k$th feature with associated variance $\tau_k^2$. $\boldsymbol{\Lambda}_k$ is the $k$th row of the $K$ by $L$ matrix of factor loadings. We assume that the $\boldsymbol{\epsilon}_{k}$'s are independent of each other and also from $\mt{z}$. The shared factors, $\mt{z}$, are assumed homoscedastic, where the columns have fixed variances equal to 1 \cite{RN674,RN562}. In this work, these shared factors are used to simplify our understanding of the spatial patterns of unhealthy behaviors and represent the building blocks for the AIBIC product described later. The covariance structure of the GSCM is described in Section C of the Additional File. 

Since factor models are notorious for being unidentifiable, there is an abundance of research in this area \cite{RN681, RN673, RN680}. In this work, we use the hierarchical structural constraint \cite{RN679} where the factor loading matrix, $\boldsymbol{\Lambda}$, is constrained to be lower triangular with strictly positive diagonal entries $\lambda_{ii} > 0$ \cite{RN674}. This constraint enforces bounds on the number of free parameters in $\boldsymbol{\Lambda}$ and the maximum number of shared factors. In this work, where $K = 5$, $L$ was set to $2$, requiring the estimation of nine parameters in $\boldsymbol{\Lambda}$. 







\subsubsection{Spatial dependence}


Although the $N$-dimensional vectors $\boldsymbol{\epsilon}_{k}$ and $\mt{z}$ are independent of each other, we assume spatial dependence within each vector. Whilst previous research has used spatial priors on the loading matrix \cite{RN682, RN683}, shared factors \cite{RN674, RN671}, or feature-specific residual errors only \cite{RN564}, traditional SCMs assume spatial dependence in the shared factors and feature-specific residual errors \cite{RN670, RN611}. Although marginal spatial priors have been explored, particularly in geostatistical applications \cite{RN671, RN504, RN562}, conditional autoregressive priors \cite{RN363} are more common in SCMs \cite{RN546, RN6, RN687} and disease mapping applications in general \cite{RN8}.

In this work, we fit independent $N$-dimensional conditional spatial priors on the columns of $\mt{z}$ and the $\boldsymbol{\epsilon}_{k}$'s. To allow for a blend of spatially structured and unstructured variation, we used the Leroux prior \cite{RN366}, hereafter called the LCAR (see Section B of the Additional File). 

The LCAR represents a generalisation of two priors due to the inclusion of a spatial autocorrelation parameter ranging from 0 to 1. When the spatial autocorrelation parameter is equal to 1, the LCAR becomes the common intrinsic conditional autoregressive (ICAR) prior \cite{RN364}, which has been used in SCMs and BSFMs before \cite{RN670, RN682}. When the spatial autocorrelation parameter is equal to 0, the LCAR collapses to an unstructured independent and identically distributed (IID) normal prior. The spatial autocorrelation parameters from the shared factors and feature-specific residual errors are denoted as $\rho_l$ and $\kappa_k$, respectively. 

The GSCM represents several generic models as special cases. The conventional (non-measurement error) SCM is recovered when uncertainty of the input data is ignored (e.g., setting $ \sigma_{nk} = 0, \forall n,k $) and $L = 1$. Continuing to ignore measurement error, a BSFM is recovered similar to that of Nethery \etal \cite{RN674} when all $\kappa_k = 0$ or that of Mezzetti \etal \cite{RN564} when all $\rho_k = 0$. Furthermore, when all $\rho_k = 0, \kappa_k = 0$ the GSCM becomes a generic Bayesian factor model. 

\subsubsection{Priors}

To complete the Bayesian GSCM, prior distributions are chosen for the parameters $\boldsymbol{\Lambda}, \boldsymbol{\tau}, \boldsymbol{\rho}, \boldsymbol{\kappa}$. We use weakly informative $\jdist{N}{0,1}$ priors on the free parameters of $\boldsymbol{\Lambda}$. Mildly informative $\jdist{Gamma}{2,3}$ priors, which have a density of 0.04 below 0.1, were used for $\boldsymbol{\tau} = \lb{\tau_1, \dots, \tau_5}$. Mildly informative $\jdist{Beta}{6,2}$ priors, which have a density of 0.94 above 0.5, were used for $\boldsymbol{\rho} = \lb{\rho_1, \rho_2}, \boldsymbol{\kappa} = \lb{\kappa_1, \dots, \kappa_5}$. These mildly informative priors were used to improve the convergence and identifiability of the model and avoid degenerate feature-specific residual errors. A sensitivity analysis found that these priors had minimal effect on the inference of $\mt{z}$.

\subsubsection{Computation}
We used fully Bayesian inference with MCMC via the R package \texttt{rstan} Version 2.26.11 \cite{RN452}. Where possible we used the non-mean centered parameterisation for the shared factors and feature-specific residual errors. Although efficient implementations of the proper CAR \cite{donegan2021} and ICAR \cite{RN397} are available, to the best of our knowledge this is the first implementation of the LCAR in Stan. This computational novelty is widely applicable to any spatial model fitted using Stan, not just the model proposed in the current paper. Code and details can be found on Github \cite{hogg2024gscm} or in Section B.1 of the Additional File. 

We used 4000 warmup and 6000 post-warmup draws for each of the four chains. For storage reasons we thinned the final posterior draws by three, resulting in 8000 draws for inference. Convergence of the models was assessed using trace and autocorrelation plots, effective sample size, and $\hat{R}$ \cite{RN499}.

\subsubsection{Model fit metrics}
To formally compare models we used the Deviance Information Criterion (DIC), Widely Applicable Information Criterion (WAIC) \cite{RN116}, and mean absolute error (MAB) between the estimates, $\mt{Y}_{k}$, and the posterior median of the true values, $\boldsymbol{\mu}_k$. For these metrics, smaller values were preferred. 

\subsubsection{Summaries} \label{sec:summaries}

To address the first aim of this study, which was to generate a summary measure of unhealthy behaviors in Australia, the results from the GSCM are used to acquire draws for four $N$-dimensional vectors. Each vector represents the underlying scores for the four indices in the AIBIC product described later. 

The first two vectors are the columns of the shared factor scores, $\mt{z}$. The third is a weighted sum of these \cite{RN360}, where the weights, $w_1, w_2$, are based on the squared factor loadings (see the Additional File for details), via

\begin{equation*}
    w_l = \lb{ \sum_{k=1}^{5} \lambda_{kl}^2 } / \lb{ \sum_{k=1}^5 \sum_{l=1}^2 \lambda_{kl}^2}, l = 1,2.
\end{equation*}
\noindent The fourth version is obtained by weighting the third version by the SA2 population.

After acquiring these four vectors, the posterior draws were converted into posterior ranks and posterior percentiles \cite{RN687, RN560, RN360}. Posterior ranks were derived by ranking each draw, while posterior percentiles were derived by assigning each draw a percentile value within the range of 1 to 100. Although the posterior percentiles and ranks provide similar inference, the ranks provide slightly more granular information, whilst percentiles are slightly easier to interpret.

The final step to produce the AIBIC product was to appropriately summarise the draws of these vectors. We use the posterior median and 95\% highest posterior density interval (HPDI). To address the second and third aims of this study, which were to identify areas with elevated cancer risk and provide decision support to easily select appropriate areas, we calculated the posterior probabilities of the vectors exceeding the 80th, 95th, and 99th percentiles \cite{RN671, RN670}. Additionally, we derived the posterior probability of the vectors being ranked within the top 10, 20, and 100 areas.

The point estimates, HDPIs, and posterior probabilities for the vectors, percentiles, and ranks are available on Github \cite{hogg2024gscm}. These summaries are provided for each of the four indices, with percentiles and ranks computed both nationally and for each of the eight states and territories of Australia. 

\section{Results}

\subsection{Exploratory Analyses}
\begin{figure}
    \centering
    \includegraphics[width=0.8\textwidth]{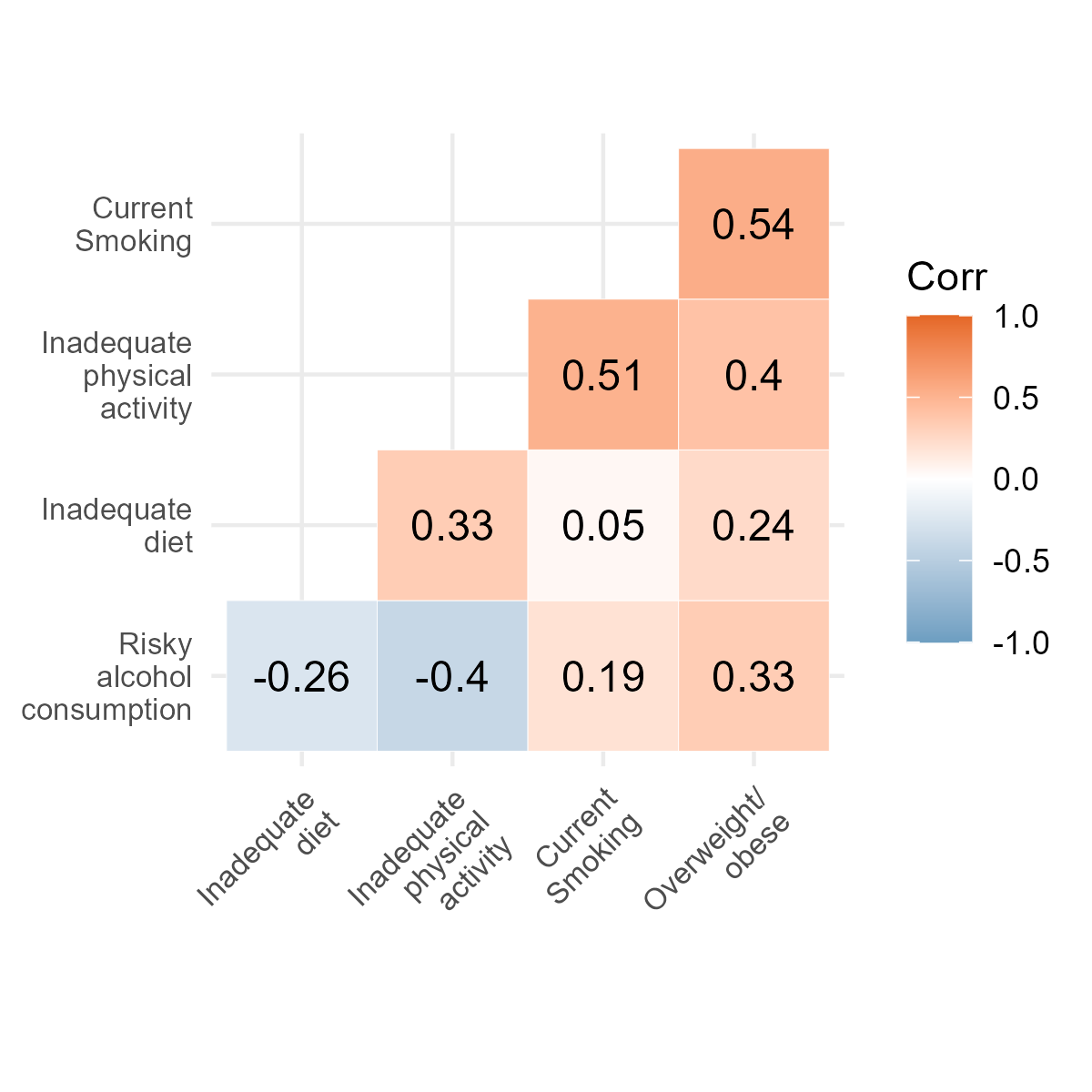}
    \caption[Visualisation of the correlation matrix for the point estimates of the five risk factors.]{Visualisation of the correlation matrix for the point estimates of the five unhealthy behaviors.}
    \label{fig:cor}
\end{figure}

\begin{figure}
    \centering
    \includegraphics[width=\textwidth]{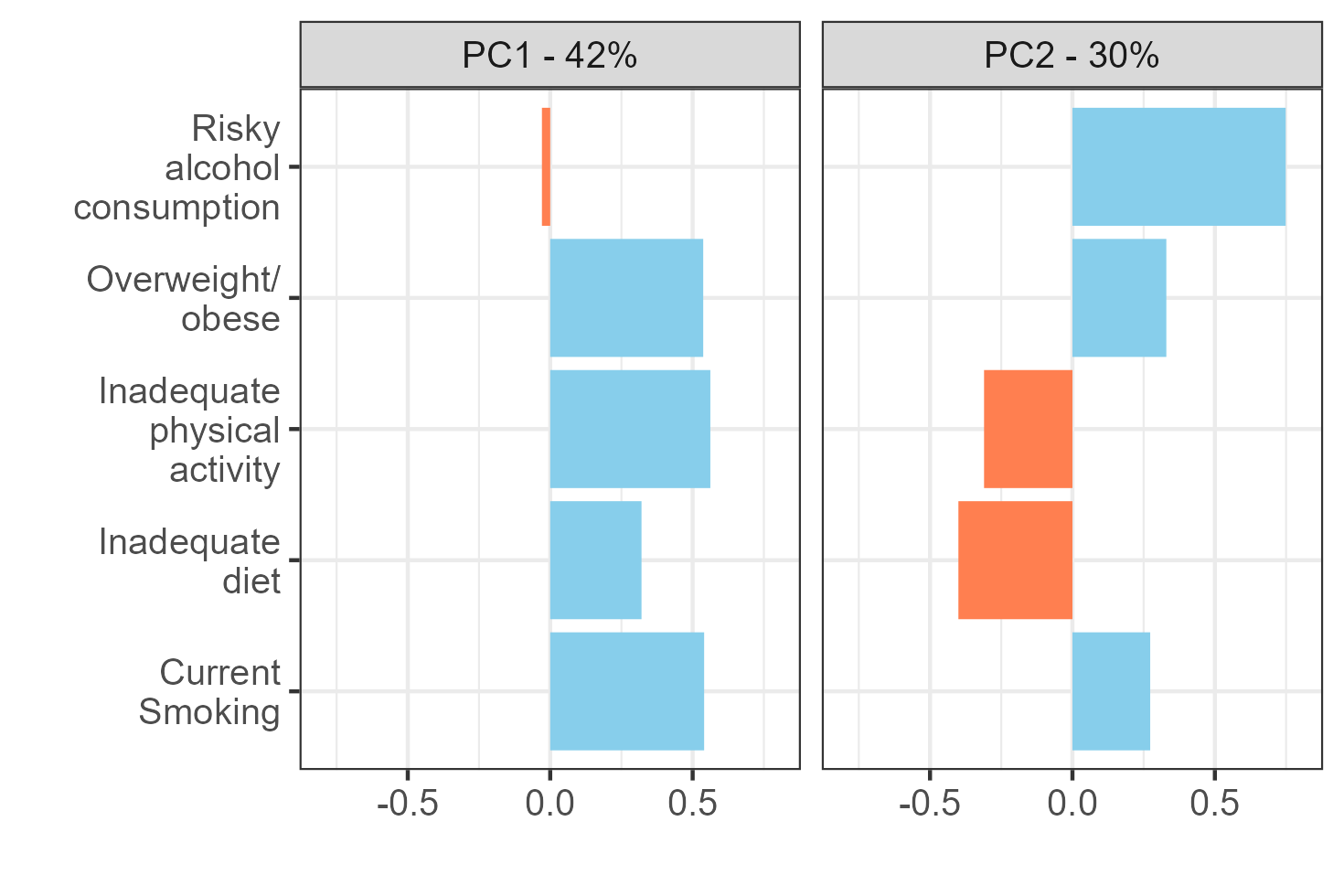}
    \caption[Visualisation of the loadings from a Principal Components Analysis (PCA) applied to the point estimates for the five risk factors.]{Visualisation of the loadings from a Principal Components Analysis (PCA) applied to the point estimates for the five unhealthy behaviors. The horizontal bars are colored according to the direction of the loading with red indicating a negative loading. Above the plots, the percentage of variation explained by Principal Component 1 (PC1) and PC2 are given.}
    \label{fig:pc_loadings}
\end{figure}

This section explores the unhealthy behaviors in terms of spatial dependency and correlation structure (\Cref{fig:cor,fig:pc_loadings}). This exploratory analysis is used to inform the fitting of the GSCM in the following sections. Section A in the Additional File provides descriptive statistics for the input features. 

The correlation coefficients of each pair of features are illustrated in \Cref{fig:cor}. It indicates a strong positive relationship between current smoking, inadequate physical activity and overweight/obese, and a negative correlation between risky alcohol consumption and inadequate physical activity. Exploring the standard errors of the input data reveals that risky alcohol consumption has the lowest error (median of 0.32), while inadequate diet (median of 0.56) and overweight (median of 0.51) have the highest. Univariate Moran's $I$ tests \cite{morani} indicate that all the risk factors exhibit a high level of spatial autocorrelation.   


Initial exploratory analysis also involved a Principal Components Analysis (PCA). The first two components accounted for 72\% of the total variation (\Cref{fig:pc_loadings}). Notably, the first principal component captured less than half of the total variation (42\%) suggesting a single shared factor would be insufficient. Furthermore, risky alcohol consumption exhibited negative or weak correlations with the majority of the other unhealthy behaviors and had a very low loading with the first principal component. Thus, using a second factor allowed us to effectively capture the distinct patterns associated with risky alcohol consumption, ensuring that these patterns were not diluted within a single shared factor or lost within the residuals. 

The results from the PCA were used to inform the ordering of the features in $\boldsymbol{\Lambda}$ in the GSCM \cite{RN682}. This ordering does not impact the theoretical model or its predictions \cite{RN678, RN679}, but can influence the shared factors and consequently, the interpretation of the indices. Given that a maximum of two shared factors is possible, only the ordering for the first two features matter in this case. We used risky alcohol consumption followed by current smoking; a choice driven by the negligible loading on risky alcohol consumption with principal component 1 (PC1) and strong positive loading with PC2. Moreover, current smoking was chosen as the second feature due to its strong positive loading with both PC1 and PC2. This ordering aimed to decorrelate the two shared factors. 


\subsection{GSCM model results}

To identify the best specification, we initially fit a range of GSCMs with increasing complexity, as described in the following paragraphs. This process allowed us to assess the impact of measurement error, the number of shared factors and spatial structures, with a small section dedicated to each. Following this assessment, we select a specific GSCM specification, which serves as the basis for developing the AIBIC product. 

Note that to improve identifiability and convergence, we used a non-spatial (e.g., IID) prior for the first feature-specific residual error in all models. 

\subsubsection{Impact of measurement error} \label{sec:me}
\begin{figure}
    \centering
    \includegraphics[width=\textwidth]{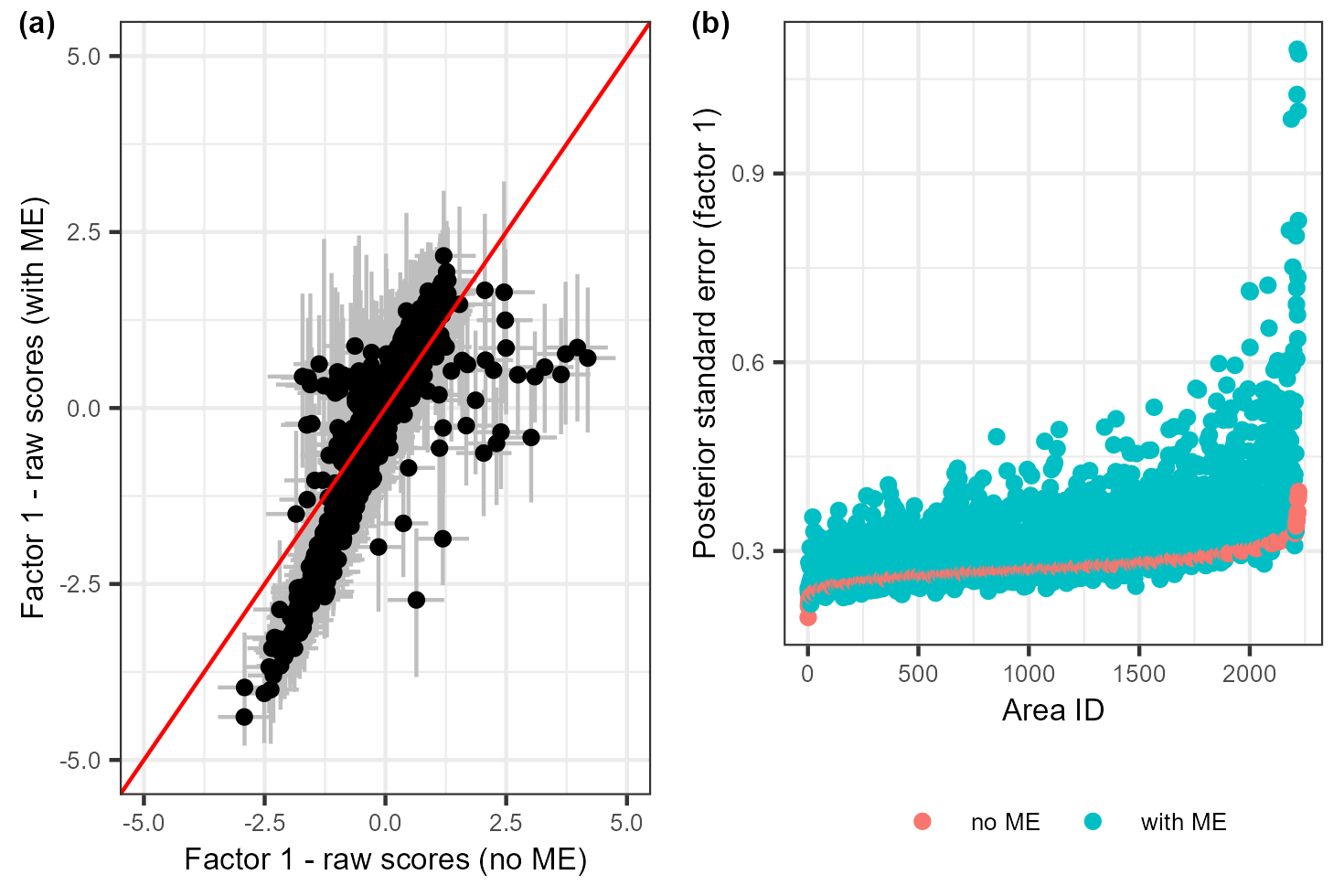}
    \caption[Comparison of the first shared factor when estimated using a two-factor GSCM.]{Comparison of the first shared factor when estimated using a two-factor GSCM with and without accommodating the measurement error (ME). Scatter plot (a) gives the posterior median and 95\% HPDIs from the GSCM with ($y$-axis) and without ($x$-axis) ME. The red diagonal line represents equivalence of the $x$ and $y$ axes. Caterpillar plot (b) compares the posterior standard deviations ranked by the ``no ME'' model.}
    \label{fig:equiv_raw_me}
\end{figure}

To assess the impact of ignoring measurement error, we compare the results from two-factor GSCMs where one model accommodates the measurement error and the other does not. Both models use LCARs for the shared factors and IID priors for the feature-specific residual errors. \Cref{fig:equiv_raw_me}, which compares the posterior median and 95\% HDPIs, confirms the expected result; the GSCM with measurement error provides shared factor 1 scores with distinct patterns (plot (a)) and considerably increased uncertainty (plot (b)). The outliers on the right of plot (a) are mostly lower socioeconomic areas within major cities. Figure S1 in Section E of the Additional File provides a similar plot for the shared factor 2 scores, illustrating slightly more agreement between the two models, but a similar increase in uncertainty.    

The factor loadings are considerably impacted when the measurement error is accommodated (\cref{table:fl_me}). The large relative changes, particularly for overweight/obese and inadequate diet, are an artefact of the higher standard errors of these input features. The largest change is for overweight/obese which loads with 0.40 for factor 1 when measurement error is ignored, but 0.07 when measurement error is accommodated. This represents both an over 5-fold decrease in central tendency and an over 4-fold increase in relative standard error.   

Note the 46\% decrease in the loading of risky alcohol consumption with factor 1 (\cref{table:fl_me}). This is noteworthy, particularly since this feature loads the strongest with this factor and has the smallest standard errors among the features, potentially explaining the marked differences in factor loadings when the measurement error is accommodated.

Considering the significant impact of measurement error, all subsequent models will accommodate the known measurement error. 

\begin{table}
\centering
\caption[Factor loadings from two-factor GSCMs.]{\small Posterior medians (95\% highest posterior density confidence intervals (HPDI)) of the factor loadings from two-factor GSCMs where one accommodates the measurement error (ME), while the other does not. Both models use LCARs for the shared factors and non-spatial priors for the feature-specific residual errors. }
\label{table:fl_me}
    \begin{tabular}{rcc|cc} 
     & \multicolumn{2}{c|}{No ME} & \multicolumn{2}{c}{With ME} \\ 
    \cline{2-5}
     & Factor 1 & Factor 2 & Factor 1 & Factor 2 \\ 
    \hline\hline
    \makecell[r]{Risky alcohol\\consumption} & 1.17 (1.07, 1.27) & 0 & 0.76 (0.73, 0.80) & 0\\
    \makecell[r]{Current\\smoking} & 0.04 (-0.04, 0.14) & 0.62 (0.57, 0.66) & -0.12 (-0.16, -0.08) & 0.54 (0.51, 0.57)\\
    \makecell[r]{Inadequate physical\\activity} & -0.77 (-0.84, -0.69) & 0.71 (0.67, 0.75) & -0.48 (-0.52, -0.43) & 0.54 (0.51, 0.58)\\
    \makecell[r]{Inadequate\\diet} & -0.30 (-0.37, -0.23) & 0.28 (0.24, 0.32) & -0.09 (-0.13, -0.06) & 0.30 (0.27, 0.32)\\
    \makecell[r]{Overweight\\/obese} & 0.40 (0.34, 0.46) & 0.68 (0.64, 0.72) & 0.07 (0.02, 0.12) & 0.66 (0.62, 0.69)\\
    \hline\hline
    \end{tabular}
\end{table}

\subsubsection{Impact of multiple shared factors}

\begin{figure}
    \centering
    \includegraphics[width=\textwidth]{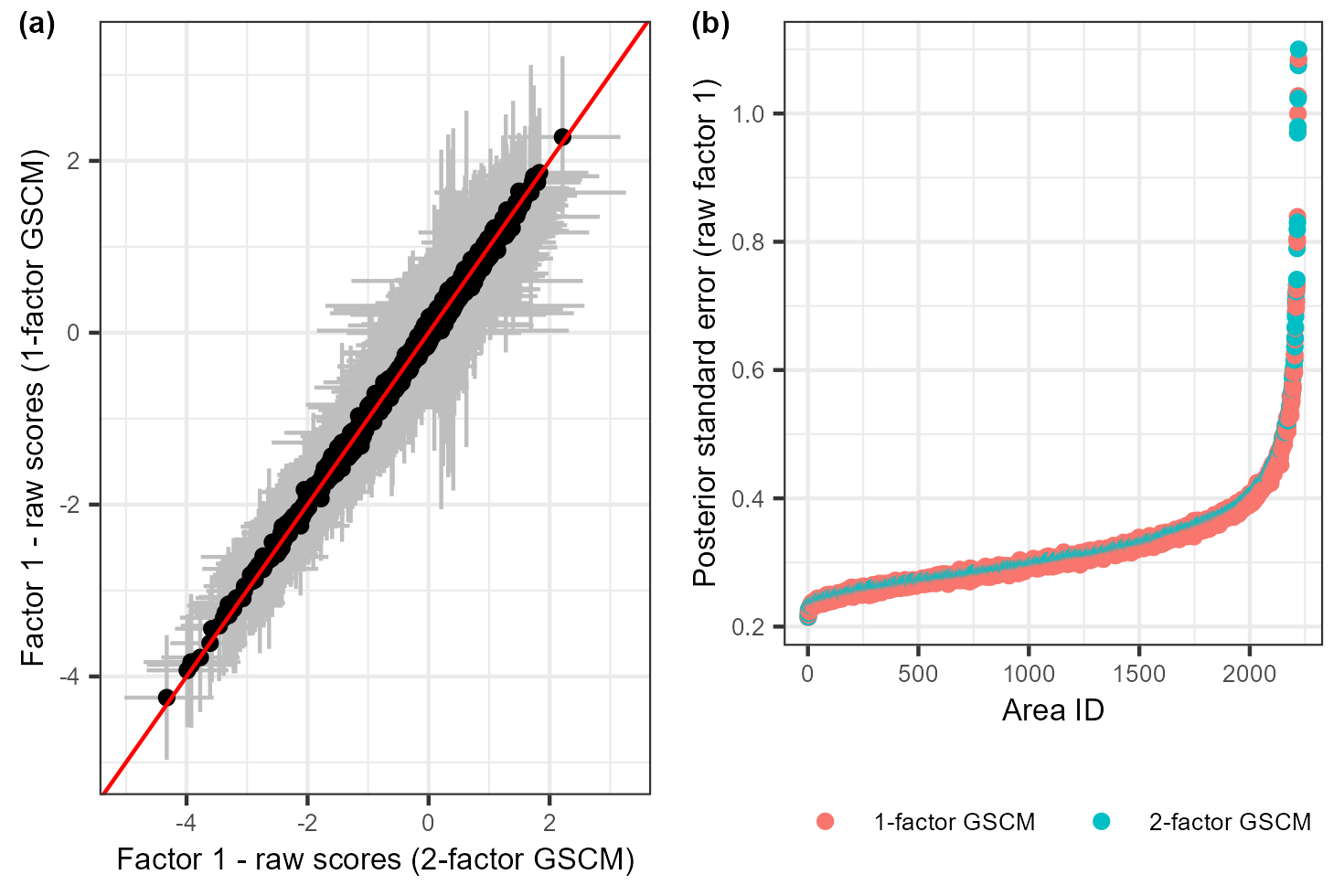}
    \caption[Comparison of the first shared factor when estimated using the 1-factor GSCM and the 2-factor GSCM.]{\small Comparison of the first shared factor when estimated using the 1-factor GSCM and the 2-factor GSCM where both have LCARs for the shared factors and feature-specific residual errors. Scatter plot (a) gives the posterior median and 95\% HPDIs from the 1-factor ($y$-axis) and the 2-factor ($x$-axis) models. The red diagonal line represents the equivalence of the $x$ and $y$ axes. Caterpillar plot (b) compares the posterior standard deviations ranked by the 2-factor model.}
    \label{fig:equiv_raw_multfactors}
\end{figure}

To assess the impact of using $L>1$ in the GSCM, we compared the first shared factor when fitting a one-factor and two-factor GSCM with LCARs for both the shared factors and feature-specific residual errors. \Cref{fig:equiv_raw_multfactors}, which compares the posterior median and 95\% HDPIs for the scores, confirms our expectation: the inclusion of the second factor in the GSCM has a negligible effect on both the central tendency and uncertainty of the factor 1 scores. 

More broadly though, the two-factor model offers a more accurate representation of the data's covariance structure, leading to the WAIC favoring it as the superior model. The WAIC for the one-factor and two-factor models are 16744 and 15591, respectively. 

Furthermore, \cref{table:fl_ml} illustrates the large loadings of all the features with the second factor, supporting its utility in this work. The factor 1 loadings are only minimally affected when factor 2 is included, with the greatest change occurring for current smoking and inadequate diet. The difficulty in interpreting factor 1, given the negative loadings, is discussed in \Cref{sec:aibic,sec:discussion}. 

\begin{table}
\centering
\caption[Factor loadings from a one-factor and two-factor GSCM.]{\small Posterior medians (95\% highest posterior density confidence intervals (HPDI)) of the factor loadings from a one-factor and two-factor GSCM with LCARs for both the shared factors and feature-specific residual errors.}
\label{table:fl_ml}
    \begin{tabular}{rc|cc} 
    \multicolumn{1}{l}{} & One-factor GSCM & \multicolumn{2}{c}{Two-factor GSCM} \\ 
    \cline{2-4}
    \multicolumn{1}{l}{} & Factor 1 & Factor 1 & Factor 2 \\ 
    \hline\hline
    \makecell[r]{Risky alcohol\\consumption} & 0.78 (0.74, 0.81) & 0.77 (0.73, 0.81) & 0\\
    \makecell[r]{Current\\smoking} & -0.20 (-0.25, -0.16) & -0.14 (-0.19, -0.10) & 0.55 (0.51, 0.58)\\
    \makecell[r]{Inadequate physical\\activity} & -0.46 (-0.51, -0.41) & -0.47 (-0.52, -0.42) & 0.51 (0.47, 0.56)\\
    \makecell[r]{Inadequate\\diet}  & -0.29 (-0.34, -0.24) & -0.23 (-0.28, -0.18) & 0.40 (0.36, 0.45)\\
    \makecell[r]{Overweight\\/obese} & 0.04 (-0.01, 0.10) & 0.08 (0.03, 0.13) & 0.63 (0.58, 0.67)\\
    \hline\hline
    \end{tabular}
\end{table}

\subsubsection{Impact of spatial priors}

We explored which combination of spatial priors provided the best fitting two-factor GSCM to these data. For the shared factors and feature-specific residual errors, three priors were possible (IID, LCAR, and ICAR) resulting in nine possible models. The results (\cref{table:model_sel}) show that the models with spatial priors for both the shared factors \emph{and} feature-specific residual errors provided a superior fit to the multivariate data.

\begin{table}
\centering
\caption[Model selection metrics for two-factor GSCM.]{\small Comparison of the two-factor GSCM when altering the prior structure for the shared factors and feature-specific residual errors. Model comparison is carried out using the Deviance Information Criterion (DIC), Widely Applicable Information Criterion (WAIC), and mean absolute error (MAB). The bolded quantities represent the lowest value in each column where appropriate.}
\label{table:model_sel}
    \begin{tabular}{llrrr}
    Shared & Specific & DIC & WAIC & MAB \\
    \hline\hline
    LCAR & ICAR & 16457 & \textbf{15583} & 0.329\\
    ICAR & ICAR & \textbf{16443} & 15589 & 0.330\\
    LCAR & LCAR & 16500 & 15591 & 0.327\\
    ICAR & LCAR & 16489 & 15601 & 0.328\\
    IID & ICAR & 17740 & 16427 & 0.295\\
    IID & LCAR & 17827 & 16476 & \textbf{0.293}\\
    LCAR & IID & 18209 & 17987 & 0.388\\
    ICAR & IID & 18201 & 17990 & 0.389\\
    IID & IID & 19634 & 19299 & 0.363\\
    \hline\hline
    \end{tabular}
\end{table}

\subsection{Area Indices of Behaviors Impacting Cancer (AIBIC)} \label{sec:aibic}

By considering the model selection criteria together (\cref{table:model_sel}), the two-factor GSCM with LCAR priors for the shared factors and ICAR priors for the feature-specific residual errors was considered the optimal model. The factor loadings are given in \cref{table:fl}, and summaries of the other model parameters are provided in Section D of the Additional File. 

\begin{table}[h]
\centering
\caption[Factor loadings from the selected GSCM.]{\small Posterior medians (and 95\% highest posterior density confidence intervals (HPDI)) of the factor loadings from the selected GSCM.}
\label{table:fl}
    \begin{tabular}{rcc}
     & Factor 1 & Factor 2\\
    \hline\hline
    \makecell[r]{Risky alcohol\\consumption} & 0.77 (0.73, 0.81) & 0\\
    \makecell[r]{Current\\smoking} & -0.15 (-0.19, -0.10) & 0.54 (0.50, 0.58)\\
    \makecell[r]{Inadequate\\physical\\activity} & -0.47 (-0.52, -0.42) & 0.51 (0.47, 0.55)\\
    \makecell[r]{Inadequate\\diet} & -0.24 (-0.29, -0.19) & 0.40 (0.35, 0.45)\\
    \makecell[r]{Overweight\\/obese} & 0.09 (0.03, 0.14) & 0.64 (0.60, 0.69)\\
    \hline\hline
    \end{tabular}
\end{table}

The shared factors from this model are used to construct the AIBIC product, which provides a relative summary measure to compare the prevalence of unhealthy behaviors between SA2s. A low index value for an area is likely to be an area where a higher proportion of the residents engage in unhealthy behaviors that can lead to cancer. 

The AIBIC product helps to address the three aims of this study: generating a summary measure of unhealthy behaviors, identifying areas with a higher prevalence of unhealthy behaviors, and providing decision support for cancer prevention in Australia. The product is tailored for broad cancer prevention strategies that aim to simultaneously address multiple unhealthy behaviors that can lead to cancer. 

In this section, we propose and compare the four AIBIC indices, each with a unique purpose and interpretation (\Cref{table:indices_names}). While the first two are derived directly from the selected GSCM, the second two are unique summations of the first two. The AIBIC product can be found on Github \cite{hogg2024gscm}. 

\begin{table}[h]
    \centering
    \caption{\small Terminology for the four indices of the AIBIC product. In the table, $z_{n1}$ is the first shared factor raw score for the $n$th area and $P_n$ is the corresponding population count.}
    \label{table:indices_names}
    \begin{tabular}{rlc}
        \hline\hline
         & Name & Source\\
         \hline\hline
        1 & Factor 1 & $z_{n1}$\\
        2 & Factor 2 & $z_{n2}$\\
        3 & \makecell[l]{Health Behaviors\\Index (HBI)}& $w_1 z_{n1} + w_2 z_{n2}$\\
        4 & \makecell[l]{Population Adjusted\\Health Behaviors\\Index (PAHBI)}& $P_n \lb{ w_1 z_{n1} + w_2 z_{n2} }$\\
        \hline\hline
    \end{tabular}
\end{table}

\subsubsection{Factor 1}

This first shared factor ($\mt{z}_1$), mapped in Figure S4 of the Additional File, could be considered an Alcohol, Activity Index as the loadings with the highest absolute magnitude appear on risky alcohol consumption and inadequate physical activity (\cref{table:fl}). The direction of the loadings suggests that areas with a high factor 1 score are areas with a high prevalence of risky alcohol consumption, but a low prevalence of inadequate physical activity. We discuss the implications of the negative loadings later (\Cref{sec:discussion}). 

\subsubsection{Factor 2}

This second shared factor ($\mt{z}_2$), mapped in Figure S5 in the Additional File, could be considered a Smoking, Activity and Weight Index, given its strong loadings with these unhealthy behaviors. Thus, factor 2 represents an approximate average of the unhealthy behaviors, excluding risky alcohol consumption. 

\subsubsection{Health Behaviors Index (HBI)}

Although factor 1 or 2 could feasibly be used to address the first and second aims of this study, justifying the sole use of either is challenging. While factor 1 can be interpreted as risky alcohol consumption minus the average of the other features, factor 2 averages all the features, except risky alcohol consumption. 

\begin{figure}[H]
    \centering
    \includegraphics[width=0.6\textwidth]{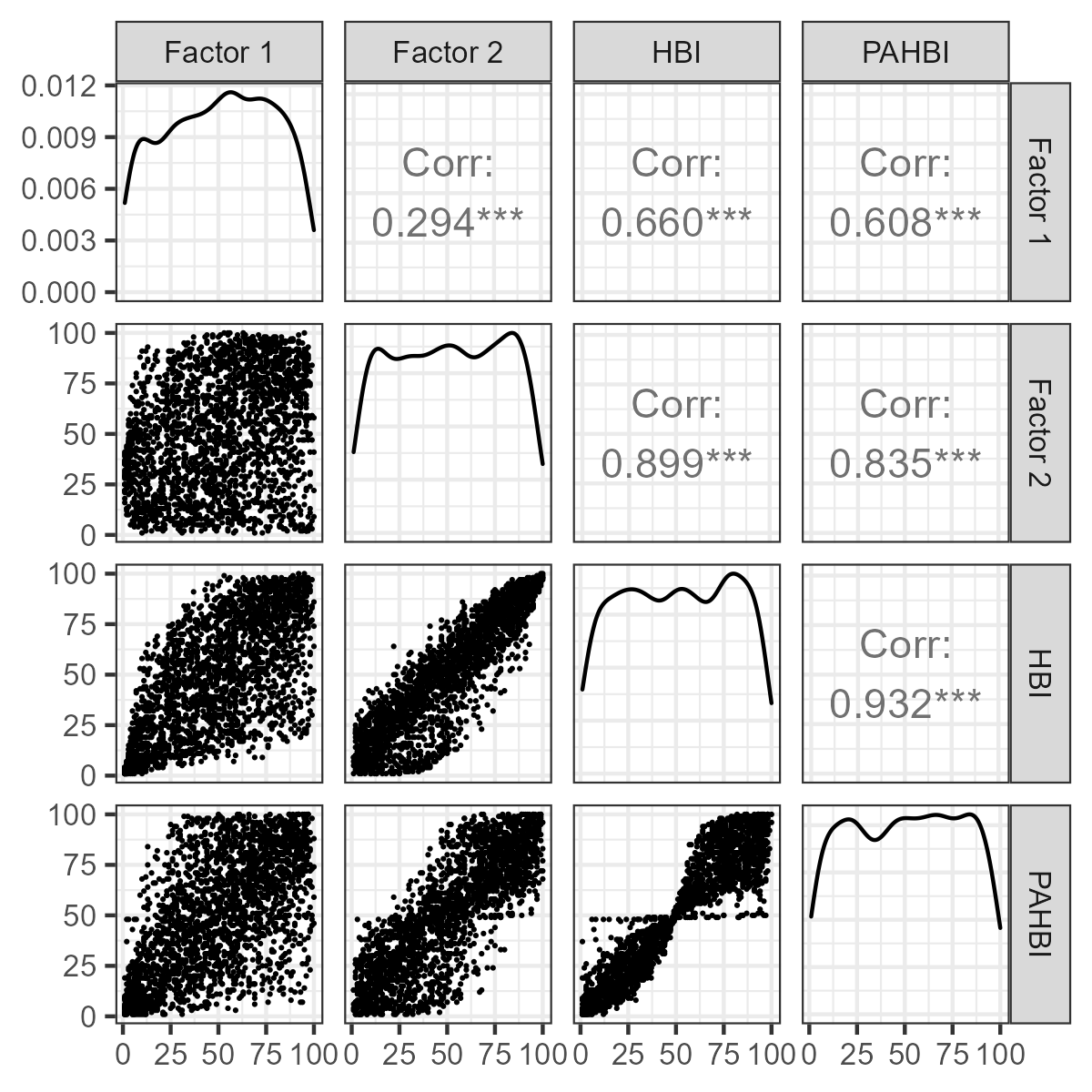}
    \caption[Pairs plot of the four indices in the AIBIC product.]{\small Pairs plot of the four indices in the AIBIC product. Posterior medians of the percentiles are used. The plots in the lower triangular squares are scatter plots, the diagonal elements are density representations and the empirical correlations between the four vectors are given in the upper triangular.}
    \label{fig:pairsperc_indexonly}
\end{figure}

Rather than select one, we followed previous advice in the Handbook of Constructing Composite Indicators \cite{RN360}, and derived a weighted average of factors 1 and 2 instead. This weighted average is the Health Behaviors Index (HBI) (\Cref{table:indices_names}). The Pearson correlations are 0.66 between factor 1 and the HBI, and 0.90 between factor 2 and the HBI (\Cref{fig:pairsperc_indexonly}); reflecting the uneven weights of 0.45 and 0.55 (\Cref{sec:summaries}). 

\begin{figure}[H]
    \centering
    \includegraphics[width=0.8\textwidth]{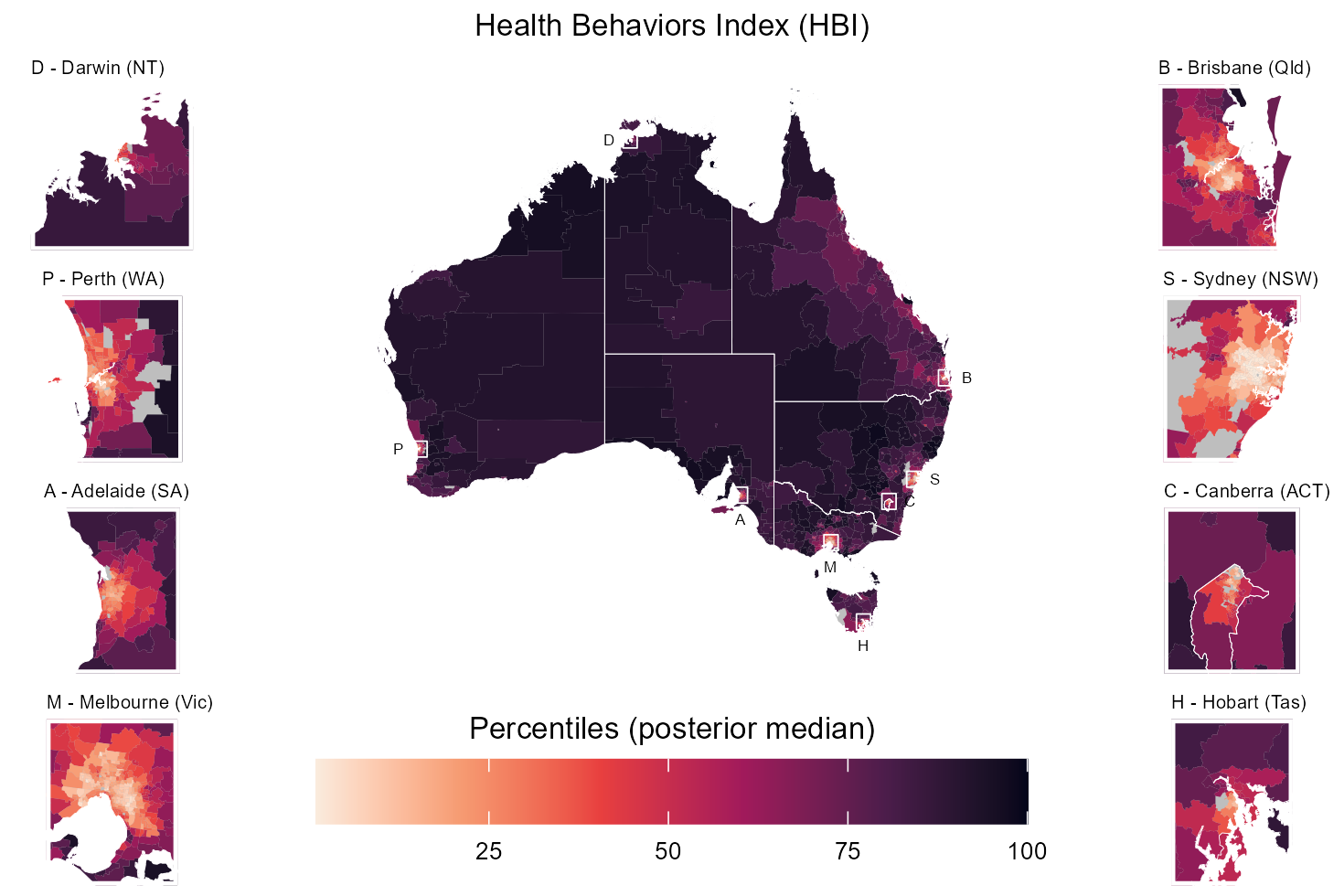}
    \caption[Choropleth maps displaying the HBI.]{\small Choropleth maps displaying the HBI, which is based on the summation of the two shared factor scores from the two-factor GSCM. The results are the posterior median of the posterior percentiles for 2221 SA2s across Australia. Higher values indicate areas with a higher prevalence of unhealthy behaviors. The map includes insets for the eight capital cities for each state and territory, with white boxes on the main map indicating the location of the inset. White lines represent the boundaries of the eight states and territories of Australia.}
    \label{fig:map_perc3}
\end{figure}

The HBI provides a single value for each area that is more representative of the five unhealthy behaviors, both in terms of point estimates and uncertainty. \Cref{fig:map_perc3} illustrates the spatial disparities in the HBI with lower values occurring in the major cities, indicating a generally lower prevalence of unhealthy behaviors in these areas. Moreover, within each remoteness category, HBI values are generally lower for less socioeconomically disadvantaged areas (Figure S6 in the Additional File). These results generally agree with the spatial patterns of the underlying unhealthy behaviors \cite{selfcite2}. 

\begin{figure}[H]
    \centering
    \includegraphics[width=0.8\textwidth]{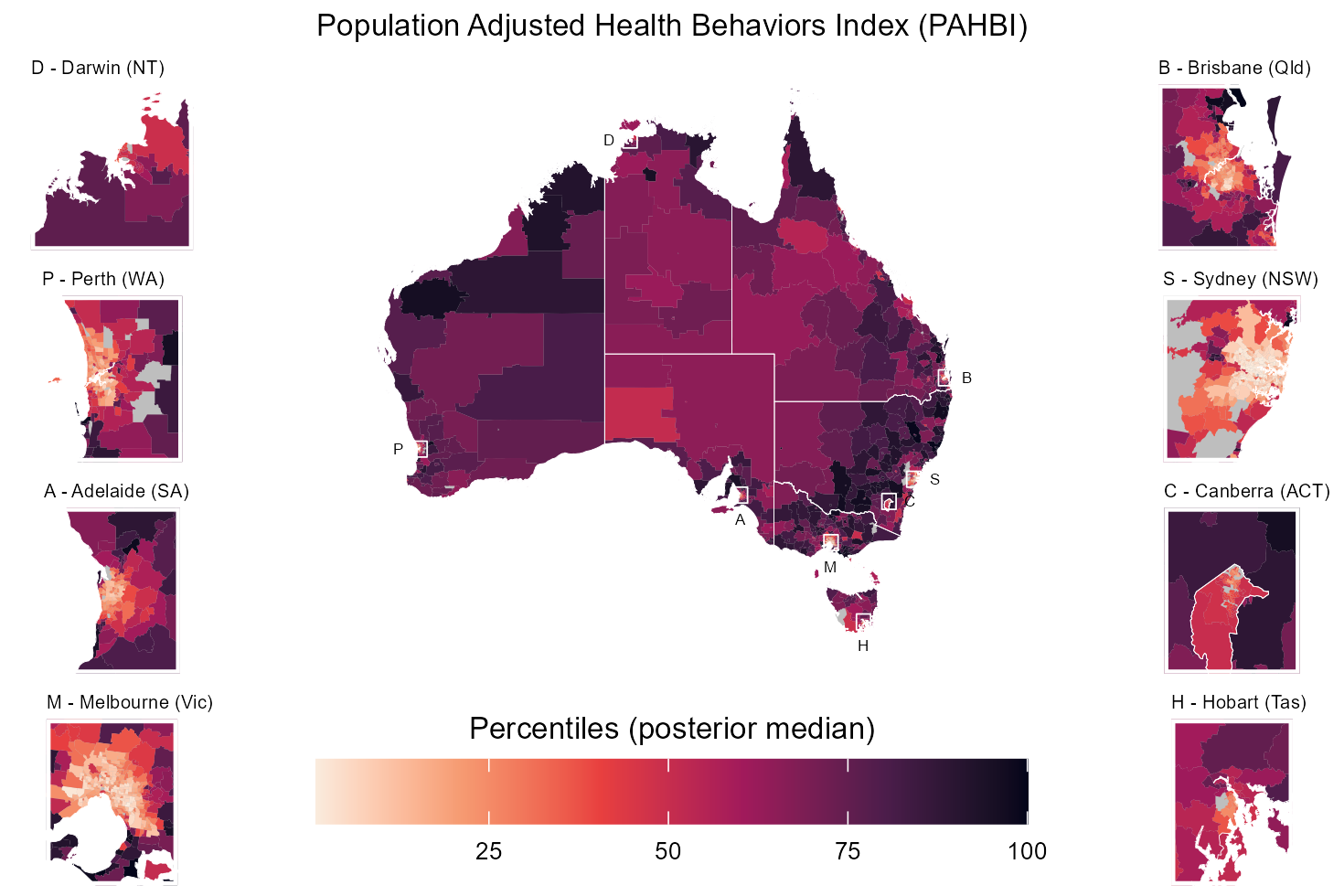}
    \caption[Choropleth maps displaying the PAHBI.]{\small Choropleth maps displaying the PAHBI, which is based on the summation of the two shared factor scores from the two-factor GSCM. The results are the posterior median of the posterior percentiles for 2221 SA2s across Australia. Higher percentiles indicate areas with a higher prevalence of unhealthy behaviors. The map includes insets for the eight capital cities for each state and territory, with white boxes on the main map indicating the location of the inset. White lines represent the boundaries of the eight states and territories of Australia.}
    \label{fig:map_perc4}
\end{figure}

\subsubsection{Population Adjusted Health Behaviors Index (PAHBI)}

Considering the third aim of this study (e.g., decision support), areas with high HBI values do not necessarily correspond with areas with large populations, where the impact may be more pronounced. To address this aim, the Population Adjusted HBI (PAHBI) is a population-weighted version of the HBI (\Cref{table:indices_names}). 

In the PAHBI, areas with high (or low) HBI values but small populations are shrunk. These areas can be seen in the PAHBI versus HBI scatter plot of \Cref{fig:pairsperc_indexonly} as the line of observations near the 50th percentile. On the other hand, areas with large populations have their index value made more extreme; ranking the area higher (or lower). 

Similar to the HBI scores, the PAHBI scores (\Cref{fig:map_perc4}) illustrate strong spatial disparities of unhealthy behaviors. Although remote areas (central Australia) do have lower PAHBI values than HBI values, even after accounting for population these regions continue to represent areas with the greatest potential for impact. 

\subsection{Comparison of indices}
The motivation for the creation of an index is to enable policymakers to prioritize health interventions in a geographically targeted manner. Figure S3 in the Additional File shows the correlations between the AIBIC indices and the individual unhealthy behaviors. 

In this section, we compare how the different indices result in unique decisions when a policymaker can only, due to feasibility or monetary constraints, select four areas to apply a new health campaign aimed at reducing unhealthy behaviors. \cref{tab:top4} provides these decisions based on the AIBIC derived from the GSCM and those from the non-model-based methods of rank sum and PCA applied to the input estimates.

For the non-model-based methods, areas were selected according to the point estimates only. For the four indices from the AIBIC, the selected areas are those with the highest posterior probability of having percentiles above the 95th. Only areas with populations greater than 99 were included. Table S5 in the Additional File provides point estimates and standard errors for the selected areas in \cref{tab:top4}. 



While the non-model-based methods can be used to efficiently select the areas with the highest prevalence of unhealthy behaviors, these methods do not accommodate the area-specific uncertainty or population size (\cref{tab:top4}). Considering PC1 it selects four areas with very high current smoking and inadequate activity rates, with the standard errors of the estimates for these unhealthy behaviors exceeding the 98th percentile of the standard errors. Similarly, PC2 selects four areas with very high current smoking and risky alcohol consumption rates, coupled with small populations (some under 200). The standard errors for these areas are all above the 97th percentile of the standard errors. 


Considering the AIBIC, specifically the PAHBI, the standard errors for the unhealthy behaviors of the areas selected are all below the 92nd percentile (median 70th) of the standard errors. The PAHBI helps to target over 71100 people, compared to around 11300 and 5770 with PC1 and PC2, respectively. Note that the selection of which index to use is contingent on the specific requirements and context of the use case.

\begin{landscape}
\thispagestyle{empty}

\begin{longtable}{l||llr||rrrrr||rrrrr||}
    \caption[Comparison of the selected top four areas when using a variety of indices.]{\small Comparison of the selected top four areas when using a variety of indices. Only areas with populations greater than 99 were considered. For rank sum, PC1 and PC2, the four areas with the highest value are chosen. For the AIBIC indices, the selected areas have the highest posterior probability of having percentiles above the 95th. Areas that are \emph{italicised} are also in the top four areas according to their posterior probability of also being ranked in the top 10 areas across Australia for the same index. For each selected area and unhealthy behavior, we provide the actual proportion estimates (and standard errors) as percentages on the left and the nationwide percentiles of the same proportion estimates (and standard errors) on the right. A percentile of 100 should be interpreted as a very high prevalence estimate (or high standard error) of that unhealthy behavior across Australia.}\\
    \label{tab:top4}
    & & & & \rotatebox{90}{\makecell[l]{Inadequate\\physical\\activity}} & \rotatebox{90}{\makecell[l]{Risky\\alcohol\\consumption}} & \rotatebox{90}{\makecell[l]{Inadequate\\diet}} & \rotatebox{90}{\makecell[l]{Overweight\\/obese}} & \rotatebox{90}{\makecell[l]{Current\\smoking}} & \rotatebox{90}{\makecell[l]{Inadequate\\physical\\activity}} & \rotatebox{90}{\makecell[l]{Risky\\alcohol\\consumption}} & \rotatebox{90}{\makecell[l]{Inadequate\\diet}} & \rotatebox{90}{\makecell[l]{Overweight\\/obese}} & \rotatebox{90}{\makecell[l]{Current\\smoking}}\\
    \hline
    Index & \makecell[l]{Area\\Name} & State & Population & \multicolumn{5}{c}{Percentages (\%)} & \multicolumn{5}{c}{Percentiles}\\
    \hline\hline
    \makecell[l]{Rank\\Sum} & Tuncurry & NSW & 5,684 & 90 (3.0) & 42 (5.7) & 49 (5.5) & 77 (4.4) & 28 (6.5) & 92 (82) & 97 (95) & 78 (94) & 99 (89) & 97 (96)\\
    & \makecell[l]{Sussex\\Inlet -\\Berrara} & NSW & 3,971 & 89 (2.7) & 41 (5.0) & 50 (4.8) & 78 (3.6) & 25 (5.2) & 89 (71) & 96 (93) & 80 (90) & 100 (79) & 95 (94)\\
    & Mannum & SA & 5,548 & 90 (2.5) & 38 (4.6) & 51 (4.4) & 76 (3.5) & 22 (4.4) & 93 (66) & 91 (91) & 86 (88) & 97 (75) & 88 (92)\\
    & \makecell[l]{Far South\\West} & QLD & 2,244 & 88 (2.4) & 36 (3.9) & 54 (3.8) & 76 (3.0) & 25 (4.0) & 85 (62) & 83 (80) & 96 (78) & 96 (51) & 95 (89)\\
    \hline
    \makecell[l]{PC1\\(42\% of\\variation)} & \makecell[l]{Palm\\Island} & QLD & 1,618 & 94 (16.2) & 33 (23.4) & 61 (22.8) & 64 (24.3) & 76 (24.0) & 100 (100) & 71 (100) & 100 (100) & 34 (100) & 100 (100)\\
    & Yarrabah & QLD & 1,776 & 95 (18.7) & 28 (25.6) & 49 (26.0) & 60 (27.6) & 80 (25.7) & 100 (100) & 41 (100) & 77 (100) & 15 (100) & 100 (100)\\
    & \makecell[l]{East\\Arnhem} & NT & 6,085 & 96 (17.7) & 26 (25.3) & 33 (24.7) & 68 (26.6) & 80 (26.1) & 100 (100) & 29 (100) & 2 (100) & 57 (100) & 100 (100)\\
    & \makecell[l]{Northern\\Peninsula} & QLD & 1,864 & 93 (9.3) & 30 (15.8) & 62 (16.0) & 67 (16.7) & 58 (19.7) & 100 (98) & 51 (99) & 100 (99) & 48 (99) & 100 (99)\\
    \hline
    \makecell[l]{PC2\\(30\% of\\variation)} & Hume & ACT & 423 & 62 (26.9) & 86 (21.0) & 14 (16.0) & 69 (22.9) & 78 (28.0) & 1 (100) & 100 (100) & 1 (99) & 62 (100) & 100 (100)\\
    & \makecell[l]{Melbourne\\Airport} & VIC & 168 & 68 (16.5) & 72 (15.3) & 24 (11.0) & 63 (15.3) & 42 (20.7) & 1 (100) & 100 (99) & 1 (98) & 29 (99) & 99 (99)\\
    & \makecell[l]{Brisbane\\Airport} & QLD & 209 & 68 (15.1) & 65 (15.2) & 25 (10.8) & 60 (14.2) & 26 (16.9) & 1 (99) & 100 (99) & 1 (98) & 16 (99) & 96 (99)\\
    & \makecell[l]{Rockbank -\\Mount\\Cottrell} & VIC & 4,970 & 82 (9.3) & 46 (13.3) & 33 (9.7) & 70 (10.7) & 44 (16.3) & 28 (98) & 99 (98) & 2 (97) & 68 (98) & 99 (99)\\
    & & & & & & & & & & & & &\\
    & & & & & & & & & & & & &\\
    & & & & & & & & & & & & &\\
    & & & & & & & & & & & & &\\
    \hline
    Factor 1 & \makecell[l]{\emph{Main}\\\emph{Beach}} & \emph{QLD} & \emph{3,751} & 76 (4.1) & 50 (4.9) & 36 (3.8) & 60 (4.5) & 14 (3.7) & 5 (93) & 99 (92) & 3 (78) & 14 (90) & 49 (83)\\
    & Barton & ACT & 1,506 & 67 (4.0) & 43 (4.0) & 49 (3.5) & 59 (3.7) & 11 (2.7) & 1 (93) & 98 (83) & 72 (71) & 13 (80) & 23 (56)\\
    & \makecell[l]{\emph{Stockton -}\\\emph{Fullerton}\\\emph{Cove}} & \emph{NSW} & \emph{6,764} & 78 (3.4) & 45 (4.2) & 39 (3.4) & 68 (3.7) & 16 (3.3) & 9 (88) & 99 (88) & 5 (67) & 56 (80) & 59 (75)\\
    & \emph{Coolangatta} & \emph{QLD} & \emph{5,738} & 76 (3.5) & 43 (4.1) & 39 (3.4) & 66 (3.6) & 15 (3.3) & 6 (89) & 97 (84) & 6 (69) & 43 (78) & 55 (75)\\
    \hline
    Factor 2 & Wingham & NSW & 4,474 & 90 (2.1) & 30 (3.7) & 53 (3.8) & 76 (3.2) & 26 (4.5) & 94 (42) & 53 (76) & 93 (79) & 96 (64) & 97 (92)\\
    & \emph{Tuncurry} & \emph{NSW} & \emph{5,684} & 90 (3.0) & 42 (5.7) & 49 (5.5) & 77 (4.4) & 28 (6.5) & 92 (82) & 97 (95) & 78 (94) & 99 (89) & 97 (96)\\
    & Casino & NSW & 9,958 & 90 (2.0) & 29 (3.3) & 54 (3.5) & 75 (3.0) & 27 (3.8) & 95 (32) & 44 (63) & 96 (70) & 95 (54) & 97 (86)\\
    & \makecell[l]{\emph{Torres}\\\emph{Strait}\\\emph{Islands}} & \emph{QLD} & \emph{3,045} & 94 (10.6) & 29 (17.7) & 53 (18.4) & 67 (18.7) & 62 (21.3) & 100 (98) & 48 (100) & 93 (100) & 52 (100) & 100 (99)\\
    \hline
    HBI & \emph{Tuncurry} & \emph{NSW} & \emph{5,684} & 90 (3.0) & 42 (5.7) & 49 (5.5) & 77 (4.4) & 28 (6.5) & 92 (82) & 97 (95) & 78 (94) & 99 (89) & 97 (96)\\
    & \emph{Forster} & \emph{NSW} & \emph{12,592} & 88 (2.5) & 38 (4.2) & 49 (4.0) & 76 (3.4) & 24 (4.4) & 81 (68) & 90 (87) & 75 (82) & 97 (71) & 93 (92)\\
    & \makecell[l]{Forster-\\Tuncurry\\Region} & NSW & 5,319 & 85 (2.1) & 35 (3.3) & 49 (3.1) & 76 (2.6) & 24 (3.4) & 61 (43) & 82 (63) & 73 (55) & 97 (21) & 93 (78)\\
    & \makecell[l]{Laurieton -\\Bonny\\Hills} & NSW & 14,976 & 84 (2.8) & 39 (4.1) & 50 (3.9) & 78 (2.9) & 22 (3.7) & 51 (77) & 93 (85) & 79 (81) & 100 (49) & 89 (85)\\
    \hline
    PAHBI & \makecell[l]{\emph{Laurieton -}\\\emph{Bonny Hills}} & \emph{NSW} & \emph{14,976} & 84 (2.8) & 39 (4.1) & 50 (3.9) & 78 (2.9) & 22 (3.7) & 51 (77) & 93 (85) & 79 (81) & 100 (49) & 89 (85)\\
    & \makecell[l]{\emph{Wonthaggi -}\\\emph{Inverloch}} & \emph{VIC} & \emph{19,804} & 83 (2.5) & 37 (3.4) & 49 (3.2) & 77 (2.6) & 22 (3.1) & 38 (65) & 86 (66) & 72 (58) & 98 (19) & 88 (71)\\
    & \makecell[l]{\emph{Port}\\\emph{Macquarie -}\\\emph{East}} & \emph{NSW} & \emph{23,748} & 81 (2.7) & 39 (3.5) & 46 (3.2) & 73 (2.9) & 17 (2.9) & 24 (73) & 92 (69) & 51 (58) & 84 (46) & 65 (63)\\
    & Forster & NSW & 12,592 & 88 (2.5) & 38 (4.2) & 49 (4.0) & 76 (3.4) & 24 (4.4) & 81 (68) & 90 (87) & 75 (82) & 97 (71) & 93 (92)\\
    \hline\hline
\end{longtable}
\thispagestyle{empty}

\end{landscape}

\section{Discussion} \label{sec:discussion}

Our results indicate that for the data utilised, ignoring the heteroscedastic measurement error, relying only on a single shared factor, or accommodating spatial autocorrelation in only the shared factors or feature-specific residual errors could lead to incomplete or inefficient policy decisions. Our solution, the GSCM, is a flexible model-based approach to index creation. Bolstering its applicability, the GSCM encompasses several previous models \cite{RN674, RN564} used for index creation as special cases. With the computational benefits of Stan \cite{RN452} and a fast implementation of the LCAR \cite{RN366}, the GSCM can feasibly be used to model a wide range of input data with varying degrees of spatial autocorrelation and measurement error. Although we illustrated our approach using atlas data from one data source, the GSCM can feasibly be applied to any geographical health dataset or datasets, with or without uncertainty. It would be feasible to fit a GSCM where some features have heteroscedastic measurement error whilst others remain free of such error. Moreover, our model could be easily incorporated into larger, more complex models such as spatial structural equation models \cite{RN504, RN506}. 

As anticipated, the incorporation of the heteroscedastic measurement error considerably altered the shared factor scores (\Cref{fig:equiv_raw_me}), but the inclusion of a second factor had minimal influence on the scores of the first factor (\Cref{fig:equiv_raw_multfactors}). However, the latter comparison was made using a GSCM with spatial shared factors and spatial feature-specific residual errors. If instead, IID priors were used for the feature-specific residual errors, the shared factor 1 scores exhibited very different patterns, with the one-factor model producing scores with higher uncertainty (see Figure S2 in the Additional File). This result supports the use of traditional SCMs (which assume spatial dependence in both the shared and feature-specific parts) over BSFMs (which assume spatial dependence in only one of these). By allowing for spatial dependence to be captured by a greater number of model terms, our results became more robust. Further, GSCMs with IID feature-specific residual errors, akin to Nethery \etal \cite{RN674} BSFM, underperformed irrespective of whether the shared factors were spatial (\cref{table:model_sel}). Future research could investigate whether the BSFM could be improved by using alternative spatial priors like the BYM \cite{RN363} and BYM2 \cite{riebler2016bym2}. Consistent with previous research \cite{RN674, RN687}, the worst-performing model was the standard Bayesian factor model which excluded spatial dependence entirely (\cref{table:model_sel}).

Given the negative loadings in factor 1 (\Cref{table:fl}), interpreting the overall cancer risk profile based on factor 1 or factor 2 alone posed a challenge \cite{RN560}. This partly motivated the combined indices. Fortunately, the negative loadings were anticipated based on previous work and our exploratory analysis of the features (\Cref{fig:cor}). Previous work indicates an inverse relationship between alcohol consumption and area level socioeconomic disadvantage \cite{RN514,RN677,RN113}. 


The three aims of this work were met by creating the Area Indices of Behaviors Impacting Cancer (AIBIC). The AIBIC, developed from the output of the GSCM, provides four indices designed to help identify areas with a high (or low) prevalence of persons who engage in behaviors that can lead to cancer, including current smoking, risky alcohol consumption, overweight/obese, inadequate diet and inadequate physical activity. While factor 1 and 2 are specific to certain unhealthy behaviors, the HBI is more general and thus more suitable for future spatial regression models or analyses. The PAHBI could be used for direct policy decisions, particularly in contexts where population density is relevant. Our analyses shed light on the different policy decisions that could be made by examining indices crafted with different methods. Indeed, indices inherently involve a loss of information, rendering them less suitable for decisions relating to a single feature (e.g., current smoking only). In such cases, a direct examination of the specific feature is always preferable. 


The AIBIC (\Cref{fig:map_perc3,fig:map_perc4}) reveals a pronounced spatial autocorrelation in unhealthy behaviors and appears to reflect the remoteness \cite{RN604} and socioeconomic \cite{RN560} patterns of Australia (see Figures S6 and S7 in the Additional File). This pattern could be an artifact of the modelling \cite{selfcite2} that produced the input data for this work. Although this initially raises concerns regarding the novelty of the indices, subsequent analyses demonstrated that the AIBIC does not accurately predict the Index of Relative Socio-Economic Disadvantage from SEIFA \cite{RN560}, with the highest accuracy being a mere 16\%. On the other hand, remoteness \cite{RN604} proves to be an adequate predictor of AIBIC scores, evidenced by a maximum adjusted $R^2$ of 45\%. Thus, future research could explore the benefits or limitations of using the AIBIC, SEIFA, and remoteness variables in the same ecological model. 



Several limitations were encountered in this study. Firstly, the absence of an established area-level cancer risk index posed challenges for validation \cite{RN482}. We sought to address this by directly comparing policy decisions derived from our indices to the underlying features. Secondly, we faced challenges with convergence and identifiability in our case study, primarily due to the strong association of risky alcohol consumption with shared factor 1, which led to the feature-specific factor for this feature being weakly identified. To mitigate this, we used more informative hyperpriors and replaced the spatial prior with an IID prior. 


The third limitation is the use of only five features and two shared factors. This may limit the transferability of the GSCM to more extensive dimension reduction tasks. Our implementation of the LCAR, although efficient, still encounters computational challenges when fitting many spatial priors. Subsequent development of these models could explore the use of direct sampling \cite{white2019direct} or approximation methods such as variational inference \cite{fox_2012}.

The usefulness of an index is constrained by the quality and breadth of the input data. Unfortunately, we lacked small area level data on factors like sun exposure, a significant determinant of melanoma, and dietary factors, including red meat and fibre consumption, which are closely linked to colorectal cancer \cite{RN165, RN123}. As these data become available, future work could update the AIBIC. 



Future methodological work could continue to tailor or generalize the GSCM to include dynamic factor loadings. We experimented with using a unique factor loading matrix for each of the five remoteness categories across Australia. Although this significantly improved model fit, it made interpretation difficult and so wasn't pursued. Another avenue of future work could be the use of non-Gaussian distributions for the likelihood, such as the Beta distribution. 

Unlike Nethery \etal \cite{RN674}, we did not include population as a feature in our model. Our method of incorporating population into the PAHBI represents just one approach, with alternative methods warranting future exploration. Moreover, there is a lack of development in alternative methods of combining numerous factors into a single index \cite{RN360}. Finally, our methodology, like other factor models, is impacted by the ordering of the features, the chosen approach for ensuring identifiability and the aggregation level of the spatial data \cite{william2024}. Different choices can result in marked variability in inferences, highlighting possible avenues to improve the robustness of model-based methods of index creation. 


In conclusion, in the current paper, we have proposed a generalised shared component model that can effectively model complex correlations in multidimensional health data, which is often derived from atlas platforms such as the Australian Cancer Atlas \cite{RN26} and Social Health Atlases of Australia \cite{RN113}. The GSCM allows for multiple shared factors and heteroscedastic measurement error and provides uncertainty measures for improved decision-making. 

We applied this model to small area-level data on unhealthy behaviors that can lead to cancer, to develop the first area-level cancer risk index in Australia; the Area Indices of Behaviors Impacting Cancer (AIBIC). Successfully addressing the three aims of this study, the AIBIC product provides a relative summary measure to compare the prevalence of unhealthy behaviors between areas and is predominantly relevant for analyses and decision-making for cancer prevention. But, the AIBIC is also applicable to other disease modelling tasks and depending on the research question, could be used in conjunction with other indices such as SEIFA. Thus, not only is the proposed model a useful generalisation of a variety of commonly used model-based methods for creating indices, but the set of indices created in this work have wide applicability in public health. 

\section*{Backmatter}

\subsection*{Abbreviations}
ABS: Australian Bureau of Statistics; AIBIC: Area Indices of Behaviors Impacting Cancer; BSFM: Bayesian spatial factor model; GSCM: Generalised shared component model; HPDI: Highest posterior density interval; ICAR: Intrinsic conditional autoregressive; IID: Independent and identically distributed; LCAR: Leroux conditional autoregressive; NHMRC: National Health and Medical Research Council; PC1: Principal Component 1; PC2: Principal Component 2; PCA: Principal components analysis; SCM: Shared component model; SEIFA: Socio-Economic Indices for Areas

\subsection*{Acknowledgments}
We thank the Australian Bureau of Statistics (ABS) for designing and collecting the National Health Survey data. The views expressed in this paper are those of the authors and do not necessarily reflect the policy of QUT or CCQ.

\subsection*{Competing interests}
The authors declare that they have no competing interests.

\subsection*{Declaration of Generative AI and AI-assisted technologies in the writing process}
During the preparation of this work, the authors used ChatGPT4.0 to improve the clarity and flow of the text. After using this tool, the authors extensively reviewed and edited the content. The authors take full responsibility for the content of the publication.

\subsection*{Author's contributions}
JH: Conceptualisation, Methodology, Software, Writing-Original Draft, Writing-Review and Editing, Visualisation, Project Administration, Resources, Formal Analysis, Validation, Investigation, Data Curation. SC: Supervision, Writing-Review and Editing, Investigation. JC: Supervision, Writing-Review and Editing, Investigation. PB: Supervision, Writing-Review and Editing, Investigation. KM: Methodology, Supervision, Writing-Review and Editing, Investigation.   

\subsection*{Funding}
JH was supported by the Queensland University of Technology (QUT) Centre for Data Science and Cancer Council QLD (CCQ) Scholarship. SC receives salary and research support from a National Health and Medical Research Council Investigator Grant (\#2008313).

\subsection*{Additional File}
The Additional File provides comprehensive details on the input data, the proposed GSCM, and the efficient \texttt{stan} implementation of the LCAR prior. It also includes details on the model results and further plots visualising the AIBIC product.  

\bibliographystyle{unsrtnat}
\bibliography{ref}  

\end{document}


\maketitle

\setcounter{section}{0}
\setcounter{subsection}{0}
\renewcommand{\thesection}{\Alph{section}}
\renewcommand{\thesubsection}{\thesection.\arabic{subsection}}

\renewcommand{\thetable}{S\arabic{table}}
\renewcommand{\thefigure}{S\arabic{figure}}

\section{Data transformation}

Let $\pi_{nk}$ and $\sigma_{nk}$ be the proportion estimate and standard error for the $n$th area and $k$th risk factor from Hogg \emph{et al.} \cite{selfcite2}. To transform these to the unconstrained (log-odds) scale,

\begin{eqnarray*}
    \theta_{nk} & = & \jdist{log}{\frac{\pi_{nk}}{1 - \pi_{nk}}}
    \\
    \tilde{S}_{nk} & = & \frac{\sigma_{nk}}{ \pi_{nk} \lb{1 - \pi_{nk}}  }.
\end{eqnarray*}

Now let $\bar{\theta}_k$ and $s_k$ be the empirical mean and standard deviation of the point estimates, $\lb{\theta_{1k}, \theta_{2k}, \dots, \theta_{Nk}}$ for the $k$th risk factor. 

The last step is to mean-center and scale the estimates, $\theta_{nk}$, and standard errors, $\tilde{S}_{nk}$,

\begin{eqnarray*}
    Y_{nk} & = & \frac{\theta_{nk} - \bar{\theta}_k}{s_k} = \frac{1}{s_k} \theta_{nk} - \frac{\bar{\theta}_k}{s_k}
    \\
    S_{nk} & = & \frac{\tilde{S}_{nk}}{s_k}. 
\end{eqnarray*}

which are the input data to the GSCM proposed in the main paper. 









\begin{table}[h]
    \centering
    \caption{Descriptive statistics for the input features, $Y_{nk}$.}
    \label{tab:descriptive_point}
    \begin{tabular}{r|cccccc}
    & Minimum & 25th Quantile & Mean & Median & 75th Quantile & Maximum\\
    \hline\hline
    Risky alcohol consumption & -4.46 & -0.39 & 0 & 0.10 & 0.53 & 7.01\\
    Current smoking & -2.08 & -0.62 & 0 & -0.07 & 0.55 & 6.32\\
    Inadequate physical activity & -4.18 & -0.59 & 0 & -0.05 & 0.60 & 4.44\\
    Inadequate diet & -8.06 & -0.48 & 0 & 0.02 & 0.57 & 4.19\\
    Overweight/obese & -4.15 & -0.71 & 0 & 0.04 & 0.75 & 3.76\\
     \hline\hline
    \end{tabular}
\end{table}

\begin{table}[h]
    \centering
    \caption{Descriptive statistics for the standard deviations, $S_{nk}$, of the input features.}
    \label{tab:descriptive_sd}
    \begin{tabular}{r|cccccc}
    & Minimum & 25th Quantile & Mean & Median & 75th Quantile & Maximum\\
    \hline\hline
    Risky alcohol consumption & 0.22 & 0.28 & 0.32 & 0.38 & 4.00\\
    Current smoking & 0.28 & 0.36 & 0.41 & 0.48 & 3.49\\
    Inadequate physical activity & 0.31 & 0.41 & 0.50 & 0.63 & 12.84\\
    Inadequate diet & 0.40 & 0.49 & 0.56 & 0.68 & 6.10\\
    Overweight/obese & 0.36 & 0.44 & 0.51 & 0.61 & 4.56\\
    \bottomrule
    \end{tabular}
\end{table}

\newpage
\section{Leroux Conditional AutoRegressive prior (LCAR)} \label{supp:A}

In this section, we describe the LCAR \cite{RN366} and provide our efficient Stan implementation. Let $\mt{W}$ be a $N \times N$ binary weight matrix where $W_{n\tilde{n}} = 1$ for row $n$ and column $\tilde{n}$ if area $n$ and area $\tilde{n}$ are neighbours and zero otherwise. In addition, let $\mt{D}$ be a $N \times N$ diagonal matrix with elements $\lb{ \sum_j W_{1j}, \dots, \sum_j W_{Nj} }$ and $\rho \in [0,1)$ be the unknown SA parameter which controls the level of spatial dependence. 

For a latent vector, $\mt{z}_l$, the unit-scale LCAR prior is a $N$-dimensional multivariate normal distribution,

\begin{eqnarray}
    \mt{z}_l & \sim & \jdistu{MVN}{N}{ \mt{0}, \Sigma =  \lbs{ \mt{I}_{\text{N}} - \rho_l \mt{C} }^{-1} }.
    \\
    \mt{C} & = & \mt{I}_{\text{N}}-\mt{D}+\mt{W} \nonumber
\end{eqnarray}

where the covariance matrix is specified as such to enforce dependence between neighbouring areas and allow for local smoothing. When $\rho = 0$ the LCAR collapses to an independent and identically distributed (IID) standard normal distribution, while when $\rho = 1$, the LCAR collapses to the intrinsic CAR (ICAR) prior, 

\begin{equation}
    \mt{z}_l \sim \jdistu{MVN}{N}{ \mt{0}, \Sigma =  \lbs{ \mt{D}-\mt{W} }^{-1} }.
\end{equation}

which has a singular covariance matrix. To correct for the ICAR not being a proper probability distribution, a sum-to-zero constraint is placed on $\mt{z}_l$. Given this property, the LCAR can become unstable when $\rho \to 1$. For this reason, we restricted $p \in [0,0.99]$, which explains why the upper bounds of the 95\% HPDIs in \cref{table:sa} are 0.99, not 1.  

Another consequence of the improper nature of the ICAR is that it cannot be utilized as the likelihood for data. Thus, in our setup, we cannot fit a BSFM where the feature-specific residual errors are spatially correlated. We avoid this when fitting the GSCM as the LCAR is used as a prior distribution. 

We enact the \emph{No Island} assumption, which posits that all rows in $\mt{W}$ must have a sum greater than zero. This implies that no areas exist in isolation \cite{RN506}. To ensure the neighborhood structure was fully connected \cite{riebler2016bym2}, we treated the eastern and western SA2s at the top of Tasmania as neighbors of the furthest south SA2s in Victoria.

\newpage
\subsection{Computation}

\subsubsection{Stan code for LCAR implementation}
\begin{lstlisting}
/**
* Log probability density of the leroux conditional autoregressive (LCAR) model
* @param x vector of random effects
* @param rho spatial dependence parameter
*           MUST be strictly smaller than 1, use:
*           `real<lower=0,upper=0.99> rho;`
* @param sigma standard deviation - often set to 1
* @param C_w Sparse representation of C
* @param C_v Column indices for values in C
* @param C_u Row starting indices for values in C
* @param offD_id_C_w indices for off diagonal terms
* @param D_id_C_w indices for diagonal terms - length M
* @param C_eigenvalues eigenvalues for C
* @param N number of areas
**
@return Log probability density
**
To use: LCAR_lpdf( x | rho, sigma, C_w, C_v, C_u, offD_id_C_w, D_id_C_w, C_eigenvalues, N );
*/
real LCAR_lpdf(
    vector x,               
    real rho,                   
    real sigma,              
    vector C_w , 
    int [] C_v , 
    int [] C_u , 
    int [] offD_id_C_w ,        
    int [] D_id_C_w ,       
    vector C_eigenvalues,       
    int N                   
    ) {                 
        vector[N] ldet_C;
        vector [ num_elements(C_w) ] ImrhoC;
        vector[N] A_S;
        // Multiple off-diagonal elements by rho
        ImrhoC [ offD_id_C_w ] = - rho * C_w[ offD_id_C_w ];
        // Calculate diagonal elements of ImrhoC
        ImrhoC [ D_id_C_w ] = 1 - rho * C_w[ D_id_C_w ];
        A_S = csr_matrix_times_vector( N, N, ImrhoC, C_v, C_u, x );
        ldet_C = log1m( rho * C_eigenvalues );
        return -0.5 * ( 
        N*log( 2 * pi() ) 
        - ( N * log(1/square(sigma)) + sum( ldet_C ) ) 
        + 1/square(sigma) * dot_product(x, A_S) 
        );
}
\end{lstlisting}

\newpage
\subsubsection{R function to derive Stan elements for \texttt{LCAR\_lpdf}}
\begin{lstlisting}
#' @param W binary contiguity matrix (must be complete - no islands)
prep4LCAR <- function(W){    
    # create sparse matrices
    W <- Matrix::Matrix( W, sparse = TRUE )
    D <- Matrix::Diagonal( x = Matrix::rowSums(W) )
    I <- Matrix::Diagonal( nrow(W) )
    C <- I - D + W
    # C and W only differ by the diagonal values
    # C has -1 on off diagonals
    
    # ISSUE: Diagonal element of C is zero if area has only one neighbor
    
    # get indices for diagonals
    jt <- rstan::extract_sparse_parts( W + 5*I ) # 5 is arbritary
    # 5's will only be on the diagonals
    D_id_C_w <- which( jt$w == 5 ) 
    # any values that are not 5 are off diagonals
    offD_id_C_w <- which( jt$w == 1 )
    
    # Eigenvalues of C
    C_eigenvalues <- eigen(C)$values
    
    # get the CRS representation of C
    # add an extra 1 to all diagonals to ensure they
    # are captured by `extract_sparse_parts`
    crs <- rstan::extract_sparse_parts( C + I )
    nC_w <- length( crs$w )
    
    # Remove 1 from the diagonals 
    crs$w[D_id_C_w] <- crs$w[D_id_C_w] - 1
    
    # prepare output list
    return(
      list(C = as.matrix(C),
           C_eigenvalues = C_eigenvalues, 
           nC_w = nC_w,
           C_w = crs$w,
           C_v = crs$v,
           C_u = crs$u,
           D_id_C_w = D_id_C_w,
           offD_id_C_w = offD_id_C_w)
    )
}
\end{lstlisting}

\newpage
\section{Generalised shared component model (GSCM)}

\subsection{Implied covariance structure}

The GSCM can accommodate complex covariance structures in the multivariate spatial data. See Section 2.2 of the main paper for notational details. 

To explore the covariance structure of the GSCM, let $L = 2$ (as in the paper). The two shared factors, $\mt{z}^{(1)}, \mt{z}^{(2)}$, have the following distributions,

\begin{eqnarray*}
    \mt{z}^{(1)} & \sim & \jdistu{MVN}{N}{ \mt{0}, \Sigma\jut{sh,(1)} }
    \\
    \mt{z}^{(2)} & \sim & \jdistu{MVN}{N}{ \mt{0}, \Sigma\jut{sh,(2)} }.
\end{eqnarray*}

with unique covariance structures (i.e., $\Sigma\jut{sh,(1)}$ is the $N \times N$ covariance matrix for the shared factor 1 scores). 

Next, consider stacking the two factors ($\mt{z}^{(1)}, \mt{z}^{(2)}$) column wise into a $N \times 2$ matrix called $\mt{z}$, using the $\jdist{vec}{.}$ function. Note also that $\jdist{bdiag}{.}$ is the block-diagonal function. 

\begin{eqnarray*}
    \jdist{vec}{ \mt{z} } & \sim & \jdistu{MVN}{2N}{ \mt{0}, \jdist{bdiag}{ \Sigma\jut{sh,(1)}, \Sigma\jut{sh,(2)} } }
\end{eqnarray*}

To infer the distribution for $\mt{z} \bm{\Lambda}^T$ --- a $N \times K$ matrix --- we use a well-known vectorization rule.

\begin{eqnarray*}
    \jdist{vec}{ \mt{z} \bm{\Lambda}^T } = \lb{ \bm{\Lambda} \bigotimes \mt{I}_N } \jdist{vec}{ \mt{z} }
\end{eqnarray*}

By letting $\mt{A} = \lb{ \bm{\Lambda} \bigotimes \mt{I}_N }$ be a $KN \times 2N$ matrix, the distribution for $\mt{A} \jdist{vec}{ \mt{z} }$ is a $KN$-dimensional multivariate Gaussian distribution with zero mean and covariance matrix

\begin{eqnarray*}
    \mt{A} \left[ \jdist{bdiag}{ \Sigma\jut{sh,(1)}, \Sigma\jut{sh,(2)} } \right] \mt{A}^T 
\end{eqnarray*}

Thus, under the GSCM, $\mt{y}$ is assumed to come from a $KN$-dimensional multivariate Gaussian distribution with zero mean and covariance matrix, 

\begin{eqnarray}
    \mt{A} \left[ \jdist{bdiag}{ \Sigma\jut{sh,(1)}, \Sigma\jut{sh,(2)} } \right] \mt{A}^T + \jdist{bdiag}{ \Sigma\jut{sp,(1)}, \dots, \Sigma\jut{sp,(K)} } + \jdist{diag}{S_{11}^2, \dots, S_{NK}^2}, \label{eq:gscm_cov}
\end{eqnarray}

\noindent where $\Sigma\jut{sp,(k)}$ is the $N \times N$ covariance matrix for the normal distribution governing the $k$th feature-specific residual error, $\boldsymbol{\epsilon}_{k}$. We decompose this covariance matrix in \cref{sec:decomp}. 

Assuming for simplicity that $K = 3$\footnote{Note that this is for exposition only, it would not be identifiable to estimate two shared factors from only three features.}, we can write the first component of \eqref{eq:gscm_cov} using block matrix notation. 

\begin{eqnarray*}
    \mt{A} \left[ \jdist{bdiag}{ \Sigma\jut{sh,(1)}, \Sigma\jut{sh,(2)} } \right] \mt{A}^T & = & \begin{pmatrix}
        \lambda_{1,1} \mt{I}_N & 0\\
        \lambda_{2,1} \mt{I}_N & \lambda_{2,2} \mt{I}_N\\
        \lambda_{3,1} \mt{I}_N & \lambda_{3,2} \mt{I}_N\\
    \end{pmatrix} \times
    \begin{pmatrix}
        \Sigma\jut{sh,(1)} & 0\\
        0 & \Sigma\jut{sh,(2)}\\
    \end{pmatrix} \times
    \begin{pmatrix}
        \lambda_{1,1} \mt{I}_N & \lambda_{2,1} \mt{I}_N & \lambda_{3,1} \mt{I}_N\\
        0 & \lambda_{2,2} \mt{I}_N & \lambda_{3,2} \mt{I}_N\\
    \end{pmatrix}
    \\
    & = & \begin{pmatrix}
        \lambda_{1,1} \Sigma\jut{sh,(1)} & 0\\
        \lambda_{2,1} \Sigma\jut{sh,(1)} & \lambda_{2,2} \Sigma\jut{sh,(2)}\\
        \lambda_{3,1} \Sigma\jut{sh,(1)} & \lambda_{3,2} \Sigma\jut{sh,(2)} \\
    \end{pmatrix} \times
    \begin{pmatrix}
        \lambda_{1,1} \mt{I}_N & \lambda_{2,1} \mt{I}_N & \lambda_{3,1} \mt{I}_N\\
        0 & \lambda_{2,2} \mt{I}_N & \lambda_{3,2} \mt{I}_N\\
    \end{pmatrix}
    \\
    & = & \begin{pmatrix}
        \lambda_{1,1}^2 \Sigma\jut{sh,(1)} & \lambda_{2,1} \lambda_{1,1} \Sigma\jut{sh,(1)} & \lambda_{3,1} \lambda_{1,1} \Sigma\jut{sh,(1)}\\
        \lambda_{2,1} \lambda_{1,1} \Sigma\jut{sh,(1)} & \lambda_{2,1}^2 \Sigma\jut{sh,(1)} + \lambda_{2,2}^2 \Sigma\jut{sh,(2)} & \lambda_{2,1} \lambda_{3,1} \Sigma\jut{sh,(1)} + \lambda_{3,2} \lambda_{2,2} \Sigma\jut{sh,(2)}\\
        \lambda_{3,1} \lambda_{1,1} \Sigma\jut{sh,(1)} & \lambda_{3,1} \lambda_{2,1} \Sigma\jut{sh,(1)} + \lambda_{3,2} \lambda_{2,2} \Sigma\jut{sh,(2)} & \lambda_{3,1}^2 \Sigma\jut{sh,(1)} + \lambda_{3,2}^2 \Sigma\jut{sh,(2)}\\
    \end{pmatrix}
\end{eqnarray*}

Replacing the spatial priors (e.g. $\Sigma\jut{sh,(1)}$ and $\Sigma\jut{sh,(2)}$) with homoscedastic IID priors (e.g. $\mt{I}_N$) results in,

\begin{eqnarray*}
    \begin{pmatrix}
        \lambda_{1,1}^2 & \lambda_{2,1} \lambda_{1,1} & \lambda_{3,1} \lambda_{1,1}\\
        \lambda_{2,1} \lambda_{1,1} & \lambda_{2,1}^2 + \lambda_{2,2}^2 & \lambda_{2,1} \lambda_{3,1} + \lambda_{3,2} \lambda_{2,2}\\
        \lambda_{3,1} \lambda_{1,1} & \lambda_{3,1} \lambda_{2,1} + \lambda_{3,2} \lambda_{2,2} & \lambda_{3,1}^2 + \lambda_{3,2}^2\\
    \end{pmatrix} \bigotimes \mt{I}_N.
\end{eqnarray*} 

We can derive a similar expression when $L = 1$; the conventional SCM. 

\begin{eqnarray*}
    \mt{A} \Sigma\jut{sh,(1)} \mt{A}^T & = & \begin{pmatrix}
        \lambda_{1} \mt{I}_N\\
        \lambda_{2} \mt{I}_N\\
        \lambda_{3} \mt{I}_N\\
    \end{pmatrix} \times \Sigma\jut{sh,(1)} \times
    \begin{pmatrix}
        \lambda_{1} \mt{I}_N & \lambda_{2} \mt{I}_N & \lambda_{3} \mt{I}_N\\
    \end{pmatrix}
    \\
    & = & \begin{pmatrix}
        \lambda_{1} \Sigma\jut{sh,(1)} \\
        \lambda_{2} \Sigma\jut{sh,(1)}\\
        \lambda_{3} \Sigma\jut{sh,(1)} \\
    \end{pmatrix} \times
    \begin{pmatrix}
        \lambda_{1} \mt{I}_N & \lambda_{2} \mt{I}_N & \lambda_{3} \mt{I}_N\\
    \end{pmatrix}
    \\
    & = & \begin{pmatrix}
        \lambda_{1}^2 \Sigma\jut{sh,(1)} & \lambda_{2} \lambda_{1} \Sigma\jut{sh,(1)} & \lambda_{3} \lambda_{1} \Sigma\jut{sh,(1)}\\
        \lambda_{2} \lambda_{1} \Sigma\jut{sh,(1)} & \lambda_{2}^2 \Sigma\jut{sh,(1)} & \lambda_{2} \lambda_{3} \Sigma\jut{sh,(1)}\\
        \lambda_{3} \lambda_{1} \Sigma\jut{sh,(1)} & \lambda_{3} \lambda_{2} \Sigma\jut{sh,(1)} & \lambda_{3}^2 \Sigma\jut{sh,(1)}\\
    \end{pmatrix} 
    \\
    & = & \begin{pmatrix}
        \lambda_{1}^2 & \lambda_{2} \lambda_{1} & \lambda_{3} \lambda_{1}\\
        \lambda_{2} \lambda_{1} & \lambda_{2}^2 & \lambda_{2} \lambda_{3}\\
        \lambda_{3} \lambda_{1} & \lambda_{3} \lambda_{2} & \lambda_{3}^2\\
    \end{pmatrix} \bigotimes \Sigma\jut{sh,(1)}
    \\
    & = & \bm{\Lambda} \bm{\Lambda}^T \bigotimes \Sigma\jut{sh,(1)}
\end{eqnarray*}

The expression above is identical to Equation 12 in the work by MacNab \cite{RN546}. 

\subsection{Decomposing the variance of the input features} \label{sec:decomp}

Using the derivations provided above, in this section we provide mathematical expressions showcasing the covariance structure imposed by the GSCM when $L = 2$. 

\noindent \emph{Variance of a single value}
\begin{equation*}
    \jdist{var}{Y_{nk}} = \lambda_{k,1}^2 \Sigma\jut{sh,(1)}_{n,n} + \lambda_{k,2}^2 \Sigma\jut{sh,(2)}_{n,n} + \Sigma\jut{sp,(k)}_{n,n} + S_{nk}^2
\end{equation*}
\noindent \emph{Covariance between values in the same feature, but different areas}
\begin{equation*}
    \jdist{cov}{Y_{nk}, Y_{\tilde{n},k}} = \lambda_{k,1}^2 \Sigma\jut{sh,(1)}_{n,\tilde{n}} + \lambda_{k,2}^2 \Sigma\jut{sh,(2)}_{n,\tilde{n}} + \Sigma\jut{sp,(k)}_{n,\tilde{n}}
\end{equation*}
\noindent \emph{Covariance between values in different features but same area}
\begin{equation*}
    \jdist{cov}{Y_{nk}, Y_{n\tilde{k}}} = \lambda_{k,1} \lambda_{\tilde{k},1} \Sigma\jut{sh,(1)}_{n,n} + \lambda_{k,2} \lambda_{\tilde{k},2} \Sigma\jut{sh,(2)}_{n,n}
\end{equation*}
\noindent \emph{Covariance between values in different features and areas}
\begin{equation*}
    \jdist{cov}{Y_{nk}, Y_{\tilde{n}\tilde{k}}} = \lambda_{k,1} \lambda_{\tilde{k},1} \Sigma\jut{sh,(1)}_{n,\tilde{n}} + \lambda_{k,2} \lambda_{\tilde{k},2} \Sigma\jut{sh,(2)}_{n,\tilde{n}}
\end{equation*}

\newpage
\begin{landscape}
\section{Model results}

To derive the HBI, we take the sum of the squared loadings for each shared factor (Table 4 in the main paper). The weights are the percentage of these sums for each shared factor. The result is about 0.45 and 0.55 for shared factors 1 and 2, respectively. This approach is described in Section 6.1 (page 89) of the Handbook on Constructing Composite Indicators \cite{RN360}. 

\begin{table}[h]
\centering
\caption{\small Posterior medians and 95\% highest posterior density confidence intervals (HPDI) for the feature-specific standard deviations across the nine different two-factor GSCMs.}
    \begin{tabular}{rrccccc} 
    Shared & Specific & $\tau_1$ & $\tau_2$ & $\tau_3$ & $\tau_4$ & $\tau_5$ \\ 
    \hline\hline
    IID & IID & 0.08 (0.01, 0.17) & 0.04 (0.00, 0.09) & 0.10 (0.02, 0.17) & 0.21 (0.16, 0.26) & 0.04 (0.01, 0.09)\\
    IID & ICAR & 0.04 (0.00, 0.10) & 0.44 (0.40, 0.47) & 0.49 (0.46, 0.53) & 0.40 (0.35, 0.44) & 0.52 (0.49, 0.56)\\
    IID & LCAR & 0.04 (0.00, 0.10) & 0.46 (0.42, 0.49) & 0.51 (0.48, 0.55) & 0.42 (0.38, 0.46) & 0.54 (0.51, 0.58)\\
    ICAR & IID & 0.03 (0.00, 0.07) & 0.08 (0.02, 0.13) & 0.04 (0.00, 0.09) & 0.26 (0.21, 0.30) & 0.03 (0.00, 0.06)\\
    ICAR & ICAR & 0.03 (0.00, 0.05) & 0.27 (0.22, 0.31) & 0.22 (0.16, 0.27) & 0.39 (0.35, 0.43) & 0.21 (0.14, 0.27)\\
    ICAR & LCAR & 0.03 (0.00, 0.05) & 0.27 (0.22, 0.31) & 0.24 (0.19, 0.30) & 0.41 (0.37, 0.45) & 0.25 (0.19, 0.30)\\
    LCAR & IID & 0.03 (0.00, 0.07) & 0.07 (0.01, 0.13) & 0.04 (0.00, 0.09) & 0.25 (0.21, 0.30) & 0.03 (0.00, 0.06)\\
    LCAR & ICAR & 0.03 (0.00, 0.05) & 0.27 (0.22, 0.31) & 0.22 (0.17, 0.27) & 0.39 (0.35, 0.43) & 0.21 (0.15, 0.27)\\
    LCAR & LCAR & 0.02 (0.00, 0.05) & 0.27 (0.22, 0.31) & 0.25 (0.19, 0.30) & 0.40 (0.36, 0.45) & 0.26 (0.20, 0.31)\\
    \hline\hline
    \end{tabular}
\end{table}

\begin{table}[h]
\centering
\caption{\small Posterior medians and 95\% highest posterior density confidence intervals (HPDI) for the SA parameters in the five different two-factor GSCMs with LCAR priors. Note that the SA parameter is only used for the LCAR. $\kappa_1$ is omitted as an IID prior was used for $\boldsymbol{\epsilon}_1$ in all models.}
\label{table:sa}
    \begin{tabular}{rrcccccc}
    Shared & Specific & $\rho_1$ & $\rho_2$ & $\kappa_2$ & $\kappa_3$ & $\kappa_4$ & $\kappa_5$\\
    \hline\hline
    IID & LCAR & & & \makecell[c]{0.9893\\(0.9870, 0.99)} & \makecell[c]{0.9890\\(0.9858, 0.99)} & \makecell[c]{0.9886\\(0.9843, 0.99)} & \makecell[c]{0.9894\\(0.9876, 0.99)}\\
    \hline
    ICAR & LCAR & & & \makecell[c]{0.9846\\(0.9692, 0.99)} & \makecell[c]{0.9869\\(0.9753, 0.99)} & \makecell[c]{0.9879\\(0.9815, 0.99)} & \makecell[c]{0.9885\\(0.9829, 0.99)}\\
    \hline
    LCAR & IID & \makecell[c]{0.9870\\(0.9790, 0.99)} & \makecell[c]{0.9891\\(0.9861, 0.99)} & & & & \\
    \hline
    LCAR & ICAR & \makecell[c]{0.9868\\(0.9786, 0.99)} & \makecell[c]{0.9889\\(0.9855, 0.99)} & & & & \\
    \hline
    LCAR & LCAR & \makecell[c]{0.9868\\(0.9786, 0.99)} & \makecell[c]{0.9889\\
    (0.9856, 0.99)} & \makecell[c]{0.9848\\(0.9703, 0.99)} & \makecell[c]{0.9868\\
    (0.9751, 0.99)} & \makecell[c]{0.9879\\(0.9815, 0.99)} & \makecell[c]{0.9885\\(0.9830, 0.99)}\\
    \hline\hline
    \end{tabular}
\end{table}

\end{landscape}

\newpage
\section{Additional plots/maps}

\begin{figure}[h]
    \centering
    \includegraphics[width=\textwidth]{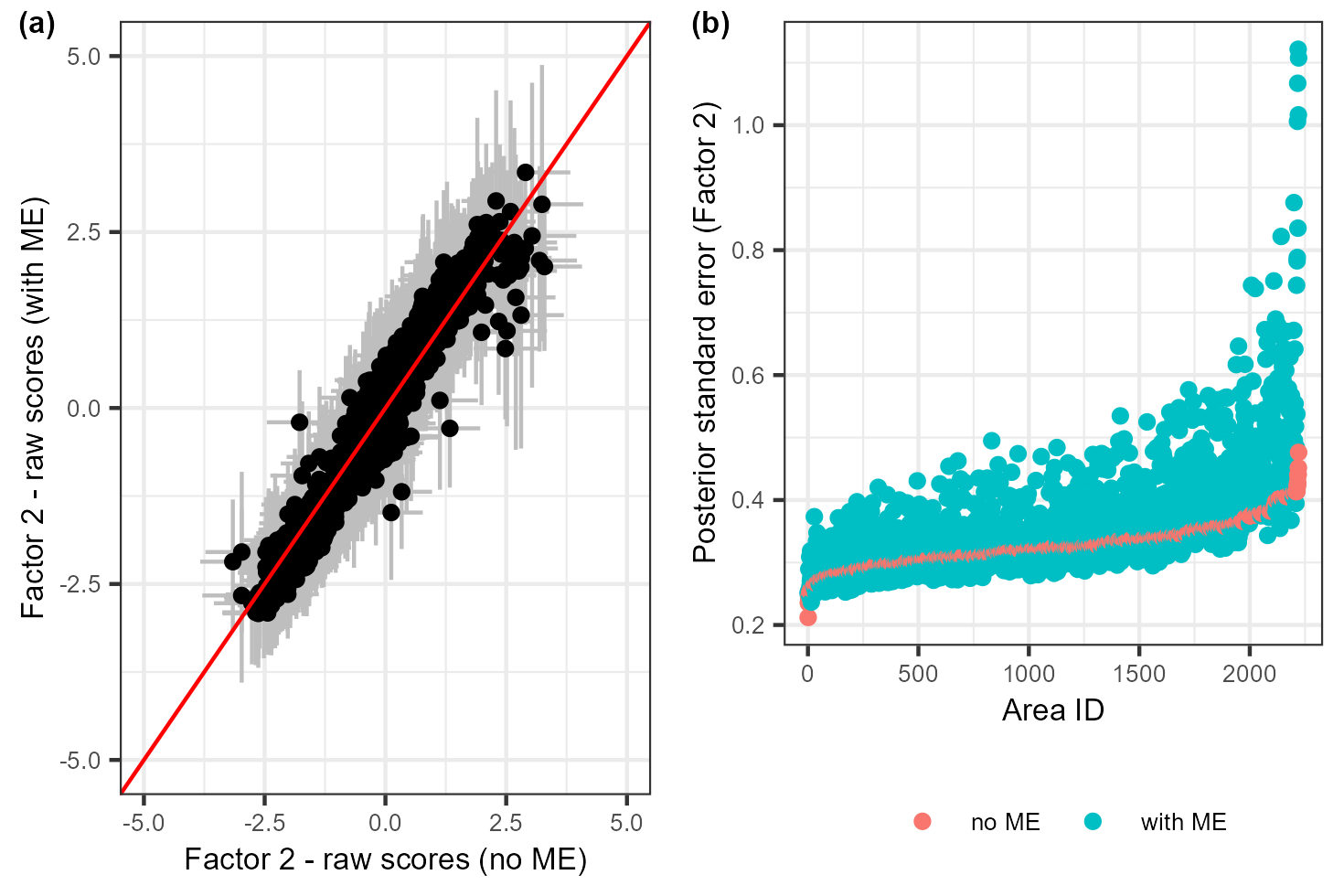}
    \caption{Comparison of the second shared factor when estimated using a two-factor GSCM with and without accommodating the measurement error (ME). Scatter plot (a) gives the posterior median and 95\% HPDIs from the GSCM with ($y$-axis) and without ($x$-axis) ME. The red diagonal line represents equivalence of the $x$ and $y$ axes. Caterpillar plot (b) compares the posterior standard deviations ranked by the ``no ME'' model.}
    \label{fig:equiv_raw_me}
\end{figure}

\begin{figure}[h]
    \centering
    \includegraphics[width=\textwidth]{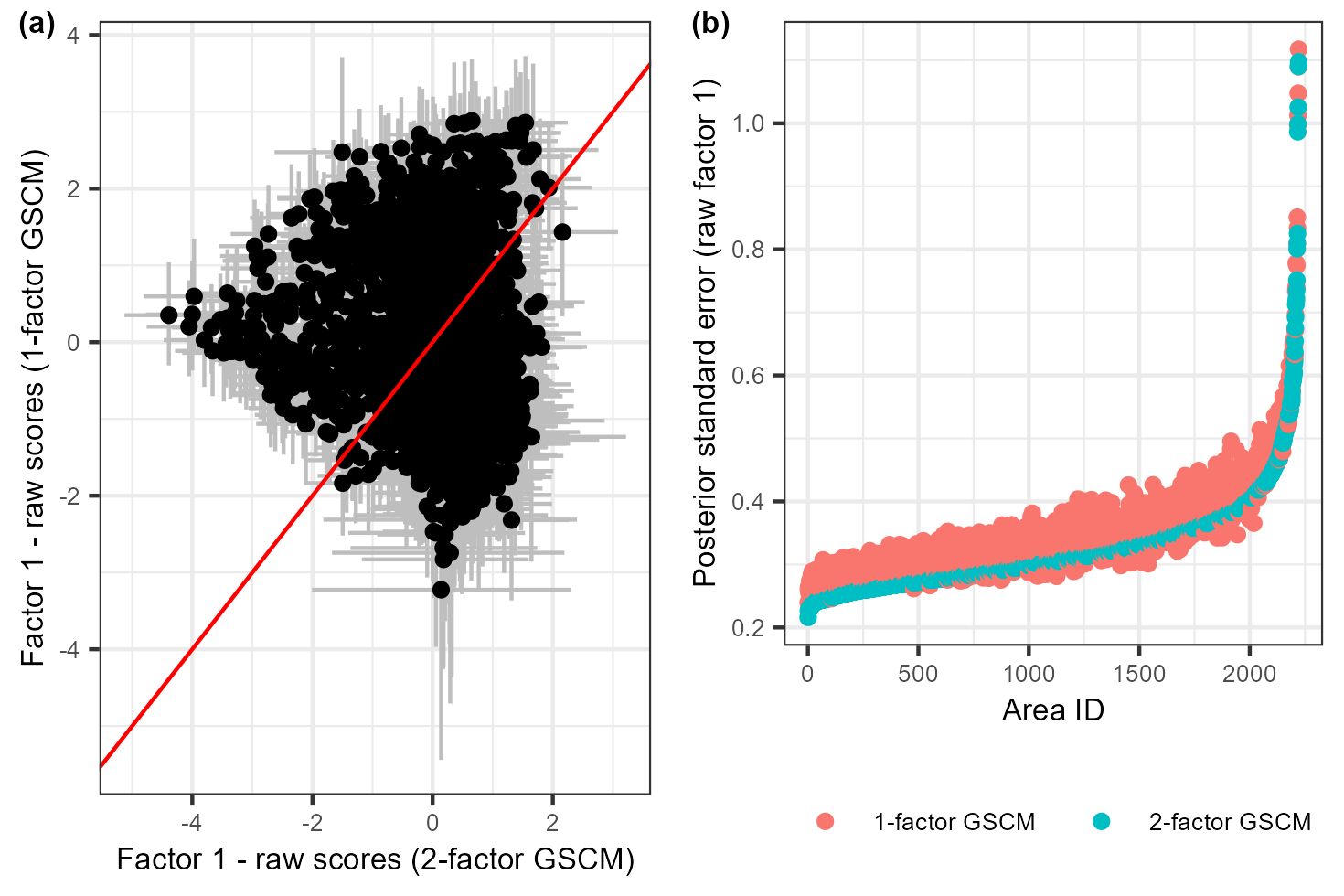}
    \caption{\small Comparison of the first shared factor when estimated using the 1-factor GSCM and the 2-factor GSCM where both have LCARs for the shared factors and non-spatial (IID) priors for the feature-specific residual errors. Scatter plot (a) gives the posterior median and 95\% HPDIs from the 1-factor ($y$-axis) and the 2-factor ($x$-axis) models. The red diagonal line represents the equivalence of the $x$ and $y$ axes. Caterpillar plot (b) compares the posterior standard deviations ranked by the 2-factor model.}
    \label{fig:equiv_raw_multfactors_IID}
\end{figure}

\begin{figure}[h]
    \centering
    \includegraphics[width=\textwidth]{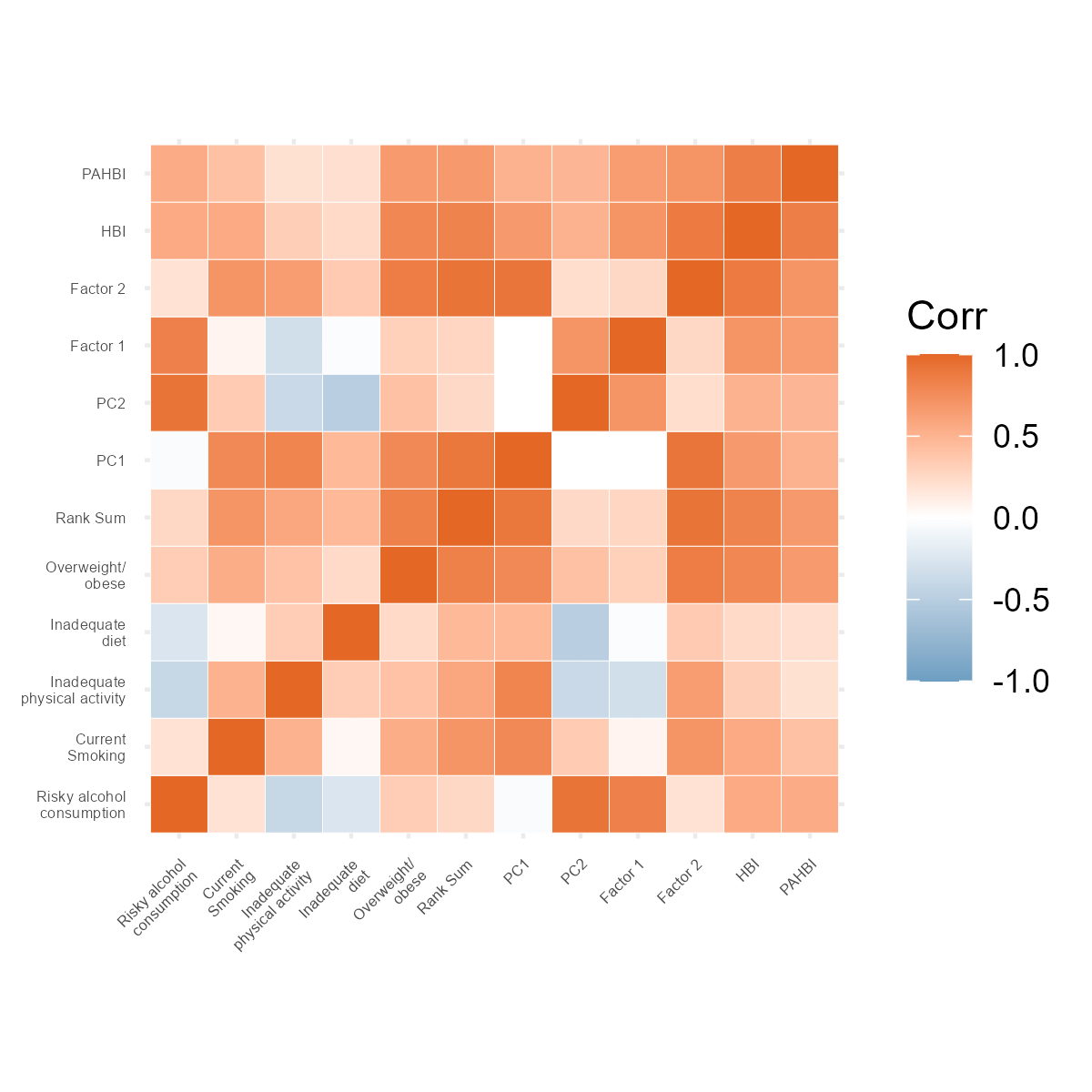}
    \caption{\small Correlation matrix of the five unhealthy behaviors, three non-model-based indices (rank-sum, PC1, and PC2), and the four indices from the AIBIC product.}
    \label{fig:cor_withfactors}
\end{figure}

\begin{landscape}

\begin{figure}
    \centering 
    \includegraphics[width=1.3\textheight]{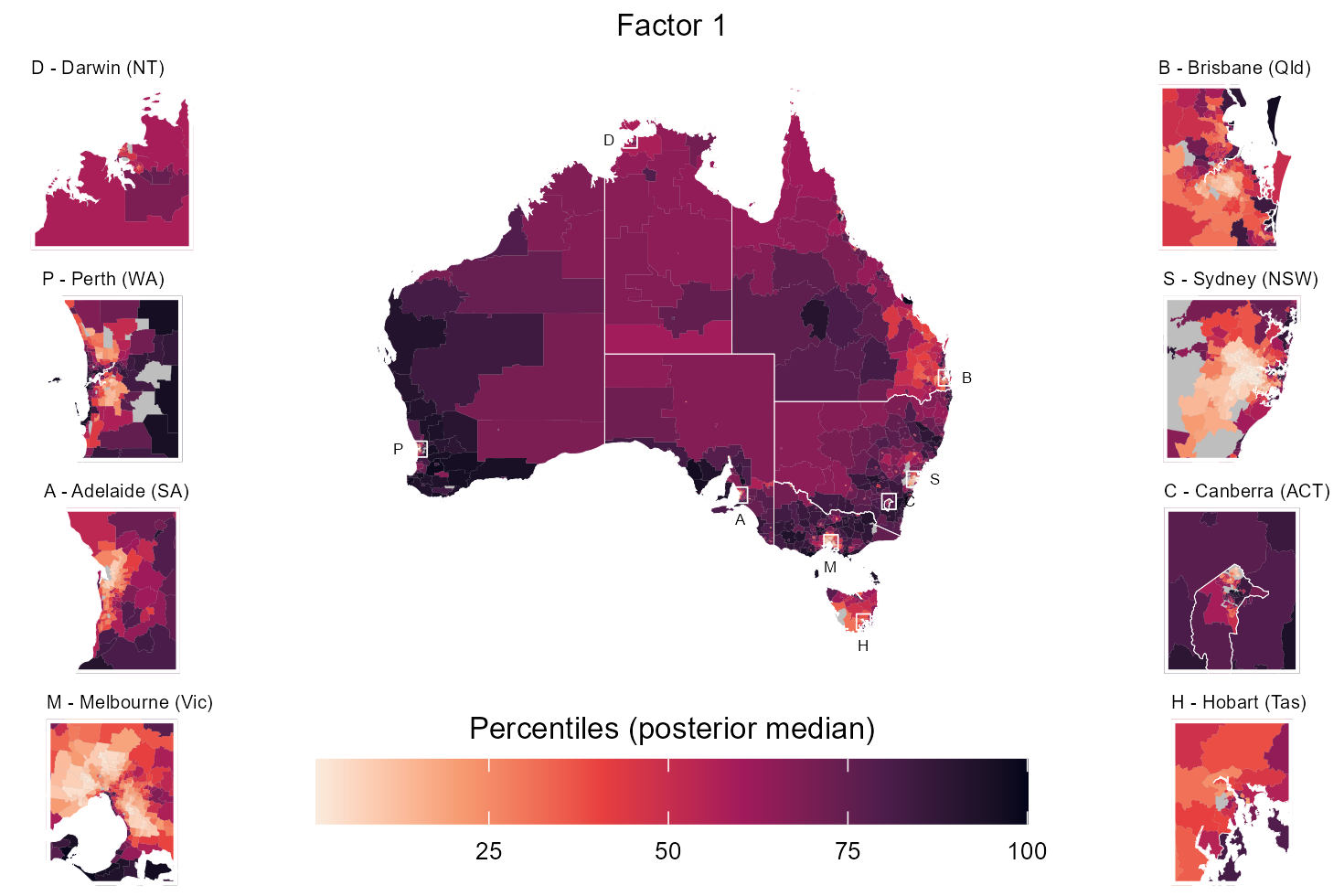}
    \caption{\small Choropleth maps displaying factor 1, which are the posterior medians of the posterior percentiles for the shared factor 1 for 2221 SA2s across Australia. Higher percentiles indicate regions with a higher prevalence of unhealthy behaviors. The map includes insets for the eight capital cities for each state and territory, with white boxes on the main map indicating the location of the inset. White lines represent the boundaries of the eight states and territories of Australia.}
    \label{fig:map_perc1}
\end{figure}

\begin{figure}
    \centering 
    \includegraphics[width=1.3\textheight]{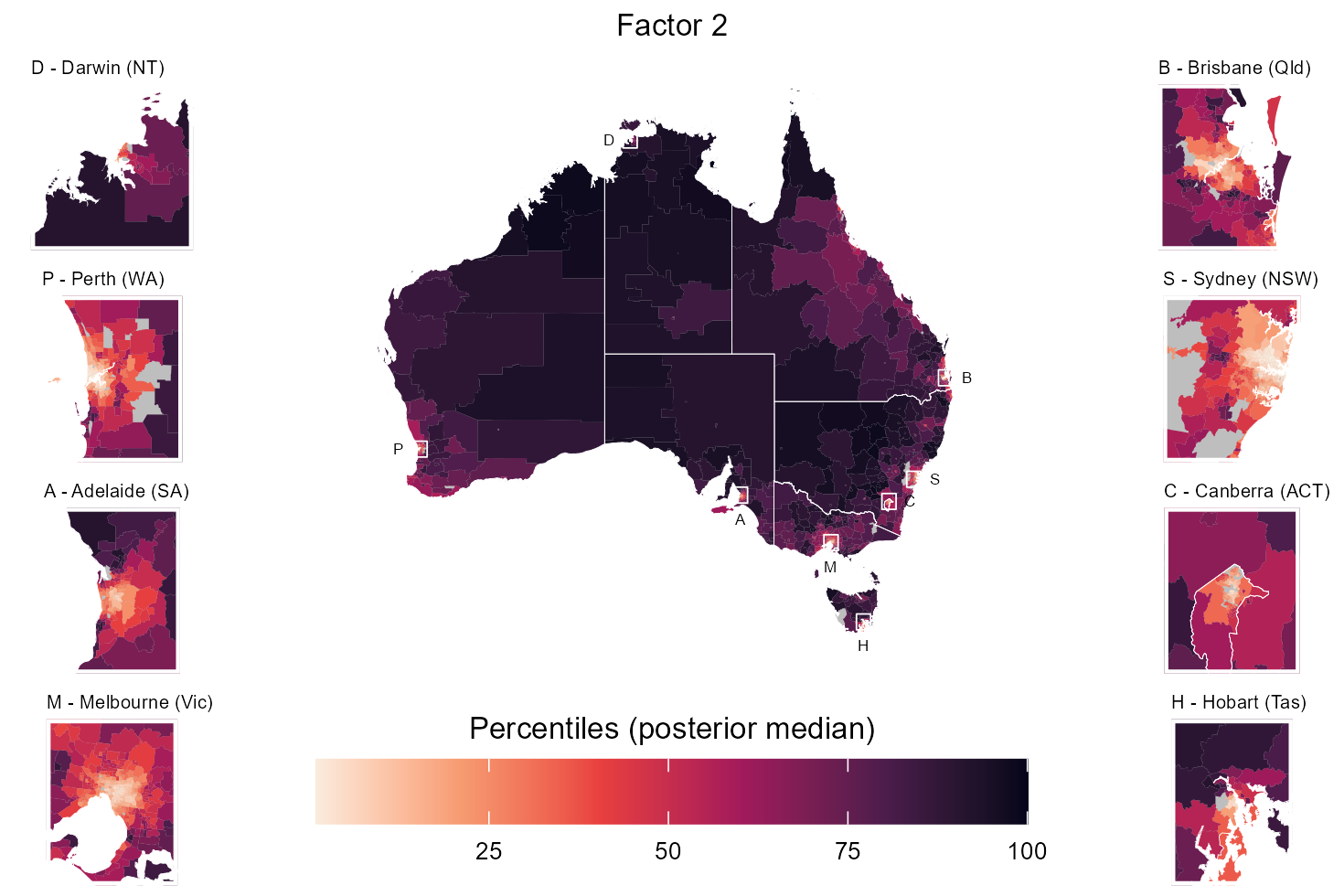}
    \caption{\small Choropleth maps displaying factor 2, which are the posterior medians of the posterior percentiles for the shared factor 2 for 2221 SA2s across Australia. Higher percentiles indicate regions with a higher prevalence of unhealthy behaviors. The map includes insets for the eight capital cities for each state and territory, with white boxes on the main map indicating the location of the inset. White lines represent the boundaries of the eight states and territories of Australia.}
    \label{fig:map_perc2}
\end{figure}

\begin{figure}
    \centering 
    \includegraphics[width=1.3\textheight]{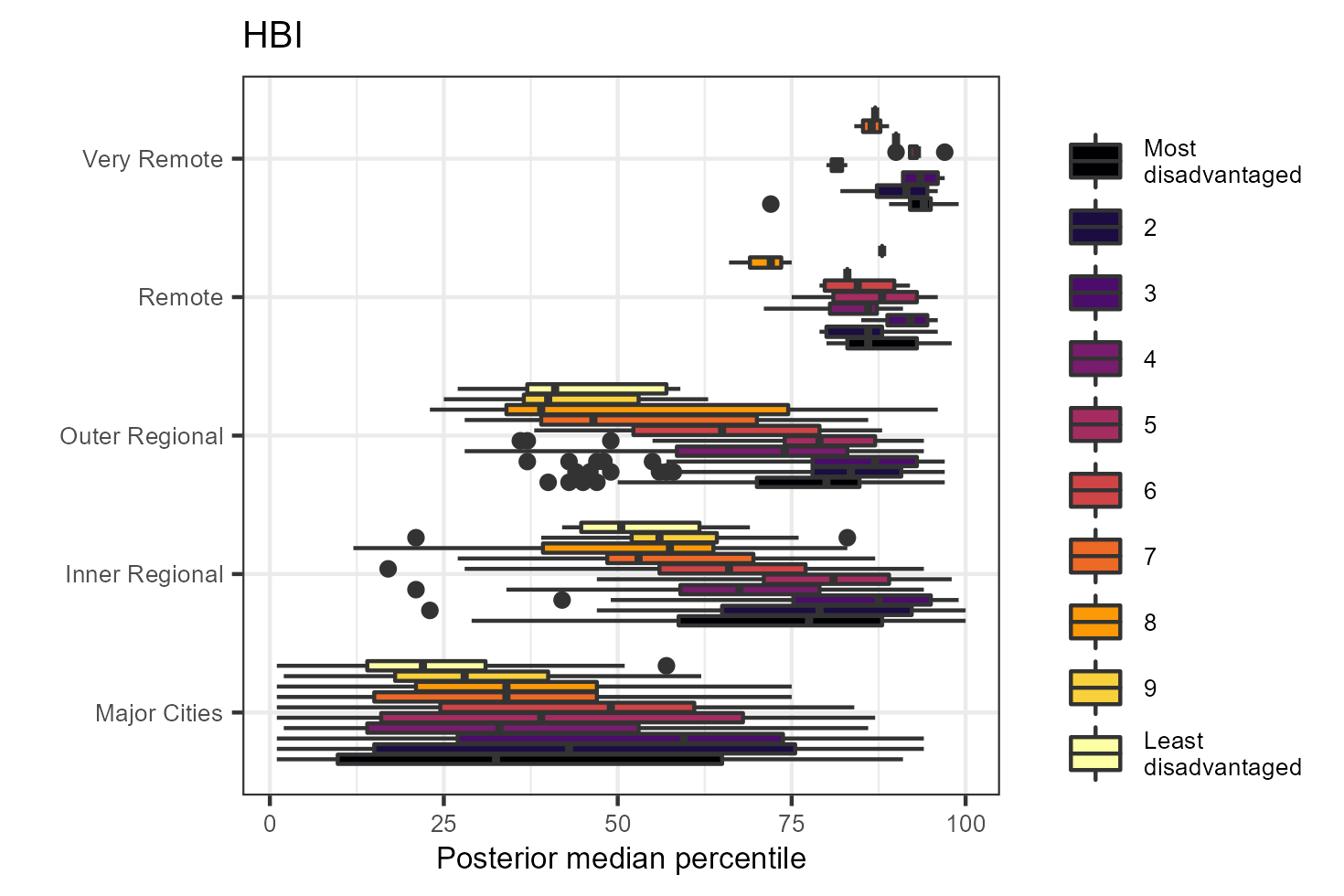}
    \caption{\small Distribution of the HBI point estimates stratified by remoteness ($y$-axis) and the Index of Relative Socio-Economic Disadvantage (IRSD) \cite{RN560}.}
    \label{fig:ra_irsd3}
\end{figure}

\begin{figure}
    \centering 
    \includegraphics[width=1.3\textheight]{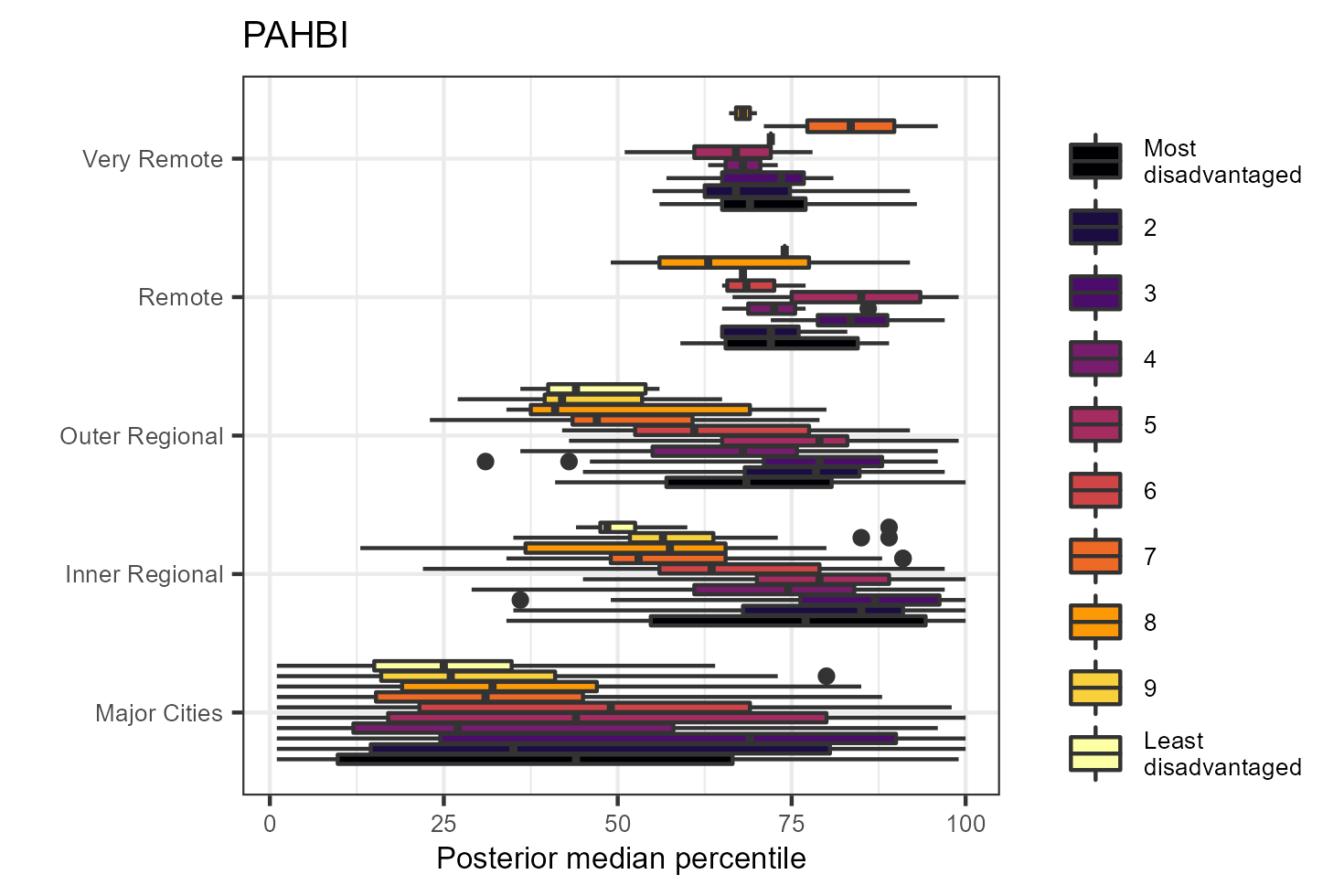}
    \caption{\small Distribution of the PAHBI point estimates stratified by remoteness ($y$-axis) and the IRSD.}
    \label{fig:ra_irsd4}
\end{figure}

\begin{figure}
    \centering 
    \includegraphics[width=1.3\textheight]{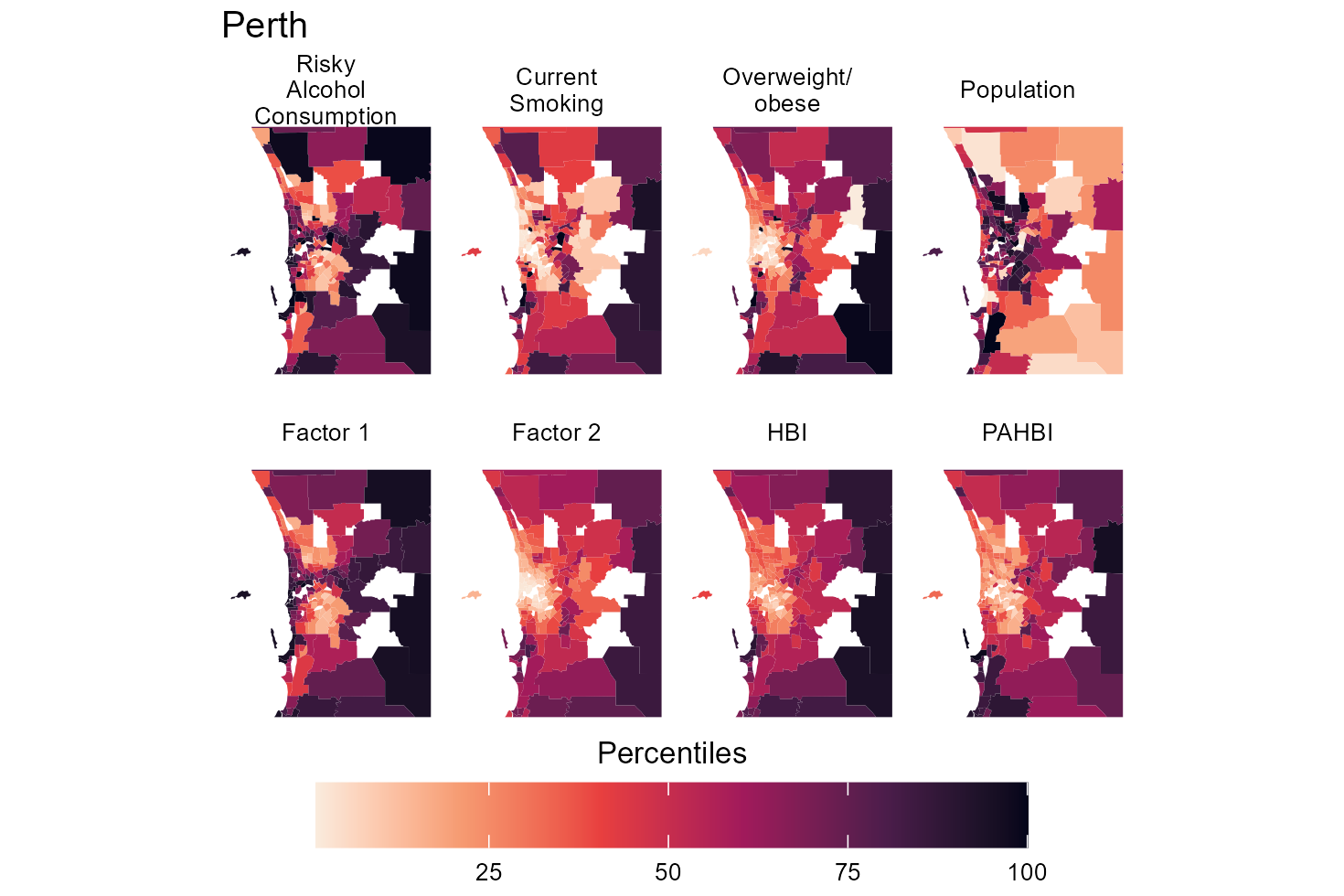}
    \caption{\small Choropleth maps of Perth, Western Australia showing percentiles for eight characteristics of this analysis. The top row displays the SA2-level prevalence of risky alcohol consumption, current smoking, overweight/obesity, and the population. The bottom row displays the four indices from the AIBIC product. AIBIC index values are posterior medians, whereas selected risk factor and population values are based on point estimates only. White areas indicate SA2s lacking data on unhealthy behaviors.}
    \label{fig:map_perc_capital_Perth}
\end{figure}

\begin{figure}
    \centering
    \includegraphics[width=1.3\textheight]{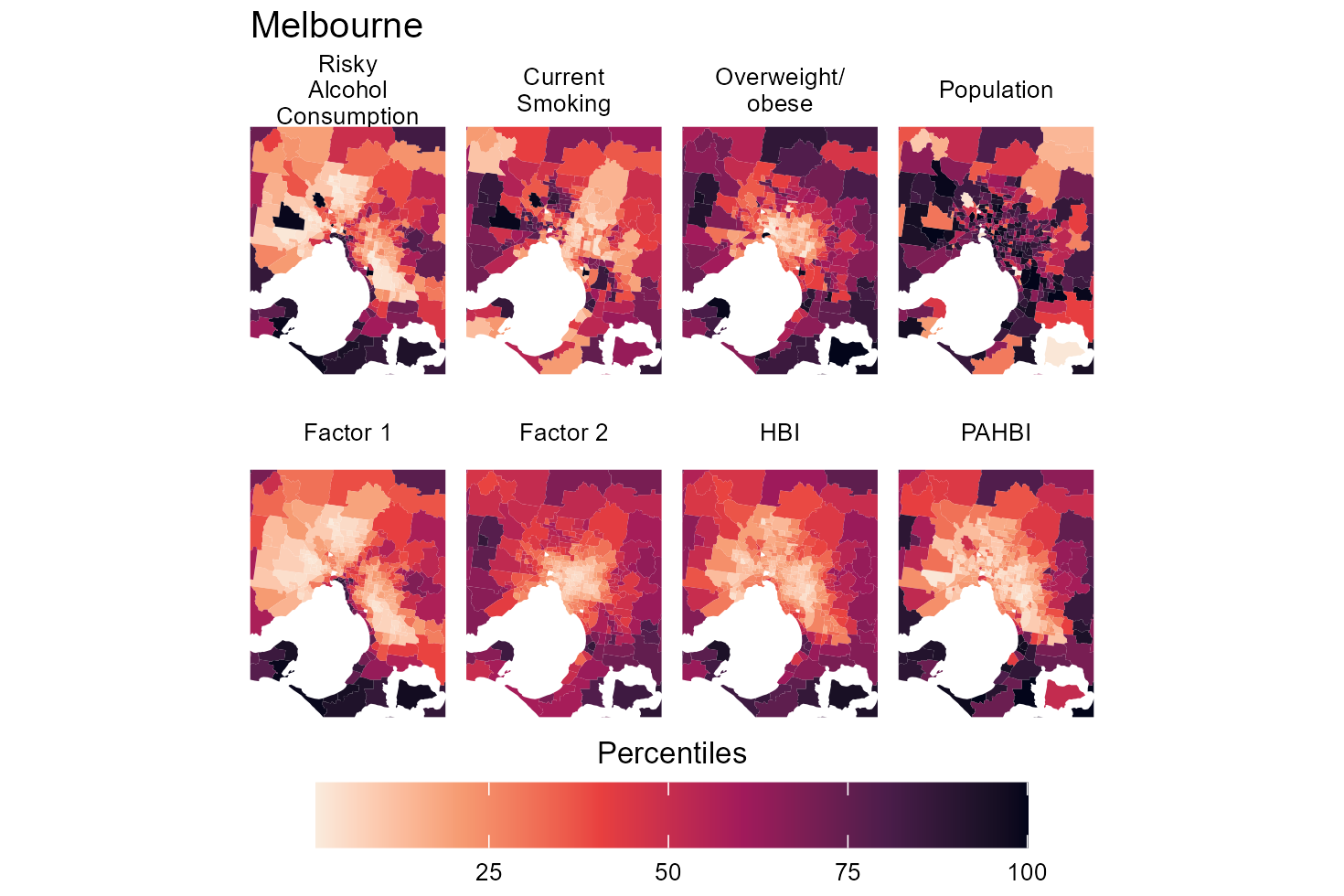}
    \caption{Choropleth maps of Melbourne, Victoria showing percentiles for eight characteristics of this analysis. The top row displays the SA2-level prevalence of risky alcohol consumption, current smoking, overweight/obesity, and the population. The bottom row displays the four indices from the AIBIC product. AIBIC index values are posterior medians, whereas selected risk factor and population values are based on point estimates only. White areas indicate SA2s lacking data on unhealthy behaviors.}
    \label{fig:map_perc_capital_Melbourne}
\end{figure}

\end{landscape}

\bibliographystyle{unsrtnat}
\bibliography{ref}  